\documentclass{article}

\usepackage{arxiv}

\usepackage[utf8]{inputenc} 
\usepackage[T1]{fontenc}    
\usepackage{hyperref}       
\usepackage{url}            
\usepackage{booktabs}       
\usepackage{amsfonts}       
\usepackage{nicefrac}       
\usepackage{graphicx}
\usepackage[numbers,sort&compress]{natbib}
\usepackage{doi}
\usepackage{amsfonts}       
\usepackage{amsmath}        
\usepackage{cleveref}       
\usepackage{microtype}      
\usepackage{multirow}
\usepackage{tabularx}
\usepackage{chngcntr}
\usepackage[final]{changes}

\title{Quantitative mapping from conventional MRI using self-supervised physics-guided deep learning: applications to a large-scale, clinically heterogeneous dataset}

\date{}

\newif\ifuniqueAffiliation

\ifuniqueAffiliation 
\else
\usepackage{authblk}

\author[1]{Jelmer van Lune}
\author[1]{Stefano Mandija}
\author[1]{Oscar van der Heide}
\author[1]{Matteo Maspero}
\author[1]{Martin B.~Schilder}
\author[2]{Jan Willem Dankbaar}
\author[1]{Cornelis A.T.~van den Berg}
\author[1]{Alessandro Sbrizzi}

\affil[1]{Computational Imaging Group for MRI Therapy \& Diagnostics, University Medical Center Utrecht, Utrecht, The Netherlands}
\affil[2]{Department of Radiology, University Medical Center Utrecht, Utrecht, The Netherlands}
\fi

\renewcommand{\shorttitle}{}
\renewcommand{\undertitle}{}
\renewcommand{\headeright}{}

\hypersetup{
pdftitle={},
pdfauthor={Jelmer van Lune},
}

\begin{document}
\maketitle

\begin{abstract}
Magnetic resonance imaging (MRI) is a cornerstone of clinical neuroimaging, yet conventional MRIs provide qualitative information heavily dependent on scanner hardware and acquisition settings. While quantitative MRI (qMRI) offers intrinsic tissue parameters, the requirement for specialized acquisition protocols and reconstruction algorithms restricts its availability and impedes large-scale biomarker research.  This study presents a self-supervised physics-guided deep learning framework to infer quantitative T1, T2, and proton-density (PD) maps directly from widely available clinical conventional T1-weighted, T2-weighted, and FLAIR MRIs. The framework was trained and evaluated on a large-scale, clinically heterogeneous dataset comprising 4,121 scan sessions acquired at our institution over six years on four different $3$~T MRI scanner systems, capturing real-world clinical variability. The framework integrates Bloch-based signal models directly into the training objective. Across more than 600 test sessions, the generated maps exhibited white matter and gray matter values consistent with literature ranges. Additionally, the generated maps showed invariance to scanner hardware and acquisition protocol groups, with inter-group coefficients of variation $\leq$~1.1\%. Subject-specific analyses demonstrated excellent voxel-wise reproducibility across scanner systems and sequence parameters, with Pearson $r$ and concordance correlation coefficients exceeding 0.82 for T1 and T2. Mean relative voxel-wise differences were low across all quantitative parameters, especially for T2 (<~6\%). These results indicate that the proposed framework can robustly transform diverse clinical conventional MRI data into quantitative maps, potentially paving the way for large-scale quantitative biomarker research. Code and model weights are available at: \url{https://github.com/JelmervanL/Quantitative-mapping-from-conventional-MRI}
	
\end{abstract}

\keywords{Quantitative MRI \and Relaxometry \and Neuroimaging \and Self-supervised deep learning \and Physics-guided neural networks \and Harmonization \and Image synthesis}


\section{Introduction}

Magnetic resonance imaging (MRI) is integral to routine clinical neuroimaging. Standard clinical protocols typically include T1-weighted (T1w), T2-weighted (T2w), and fluid-attenuated inversion recovery (FLAIR) scans to detect, characterize, and monitor neurological disorders. The signal intensity of these conventional MRIs is a function of intrinsic tissue parameters: longitudinal relaxation (T1), transverse relaxation (T2), and proton-density (PD). However, the signal intensity is also a function of extrinsic parameters such as MRI scanner hardware, vendor implementations, software version, and sequence parameters, e.g., repetition time (TR), echo time (TE), inversion time (TI), and flip angle (FA). These extrinsic parameters introduce non-physiological variability across scanner systems, protocols, and time. As a result, conventional MRI provides only relative signal intensity information and lacks absolute quantitative meaning. This limitation hinders the development of objective biomarkers, longitudinal analyses, and degrades the performance of data-driven image analysis models \citep{Hu2023,Bento2022, Mrtensson2020, Wen2023, Tixier2021}. 

Quantitative MRI (qMRI) addresses this limitation of conventional MRI by directly measuring intrinsic tissue properties such as T1, T2, and PD. This technique produces voxel-wise parametric maps with physical meaning — i.e. quantitative maps \citep{Keenan2019, Weiskopf2021}. These maps enable direct interpretation of tissue properties and comparisons across subjects, sites, and time \citep{Buonincontri2019, Gracien2020}. In contrast, such comparisons are not practical or feasible with conventional MRI. As a result, qMRI can help to develop interpretable objective biomarkers for diagnosis, disease characterization, and treatment-response monitoring \citep{Ding2021,Seiler2021,Chekhonin2024,MohamedSajer2025,Gerhalter2022,Lou2021,Springer2022}. 

Although qMRI offers clear advantages, its adoption in clinical practice remains limited. Traditional qMRI techniques require lengthy acquisitions of multiple weighted images with varying inversion or echo times, followed by exponential model fitting to estimate the quantitative parameters \citep{Serai2022}. More recently, acquisition times have been significantly reduced using dedicated specialized sequences such as MR fingerprinting (MRF) \citep{Ma2013, Choi2022}, MR-STAT \citep{Sbrizzi2018, vanderHeide2020, Liu2025}, STAGE \citep{Chen2018, Wang2018, Haacke2020}, synthetic MRI (syMRI) \citep{Hagiwara2017, Ndengera2022}, and MR multitasking \citep{Ma2020, Cao2022}. However, these dedicated sequences and their reconstruction algorithms are not widely available on clinical systems, particularly outside academic research hospitals. Furthermore, robust and validated tools for quantitative map reconstruction and biomarker extraction are limited \cite{Saltarelli2025}. As a result, quantitative maps are not widely accessible, and biomarkers developed from these maps are not yet reliable enough for broad clinical use. This creates a catch-22 situation in which limited availability hinders evidence generation, and limited evidence hinders clinical adoption \citep{Hsieh2020}.

\added{To address the limited availability of qMRI data, recent work has aimed to retrospectively generate quantitative maps from widely available conventional MRI datasets. For instance, mathematical modeling and intensity scaling techniques have attempted to estimate relaxation parameters directly from conventional scans \citep{Moskovich2024, Wiltgen2025}. However, these approaches face significant limitations. Specifically, they often require proton-density-weighted (PDw) images, which are rarely acquired in modern clinical practice \citep{Moskovich2024}, or rely on the segmentation of specific reference regions \citep{Wiltgen2025}. Furthermore, these methods are sensitive to sequence parameter variations (e.g., changes in TR or TE) and generally assume static, single-scanner protocols. As a result, they yield relative, unitless metrics rather than absolute physical values.}

\replaced{To overcome the limitations of these mathematical models, recent work has employed supervised deep learning to learn the quantitative mapping between paired conventional and quantitative reference scans \citep{Qiu2022, Moya-Sez2021, Moya-Sez2025, Wang2025, Sun2023, Wu2020, Chen2026}. However, such paired scans are rarely available at scale. Thus, this scarcity prevents us from leveraging the widely available conventional clinical MRI archives. Moreover, quantitative reference scans themselves are biased by the specific qMRI technique used, due to incomplete signal models and varying sensitivities to unmodeled parameters. As a result, there is currently no universally accepted gold standard or consensus on qMRI technique \citep{Bojorquez2017, Asslnder2025, Saltarelli2025}. These limitations restrict the scalability and robustness of supervised deep learning for quantitative mapping from conventional MRI.}{To address the limited availability of qMRI data, recent work has aimed to retrospectively generate quantitative maps from widely available conventional MRI datasets using deep learning. A typical approach employs supervised learning, where models learn the quantitative mapping between paired conventional and quantitative reference scans \mbox{\citep{Qiu2022, Moya-Sez2021, Moya-Sez2025, Wang2025, Sun2023, Wu2020, Chen2026}}. However, such paired scans are rarely available at scale. Thus, it prevents us from leveraging the widely available conventional clinical MRI archives. Moreover, quantitative reference scans themselves are biased by the specific qMRI technique used, due to incomplete signal models and varying sensitivities to unmodeled parameters. As a result, there is currently no universally accepted gold standard or consensus on qMRI technique \mbox{\citep{Bojorquez2017, Asslnder2025, Saltarelli2025}}. These limitations restrict the scalability and robustness of supervised deep learning for quantitative mapping from conventional MRI.}

A promising alternative to supervised learning is self-supervised learning. In this deep learning paradigm, training objectives are derived directly from the input data. This allows a model to learn informative representations without requiring external reference data \cite{Rani2023}. In the context of MRI, self-supervision can be established through physics-guided constraints, where MR signal models are integrated directly into the training objective. Self-supervised physics-guided models have been applied in several MRI domains, including image reconstruction \citep{Liu2021}, accelerated quantitative fitting from dedicated acquisitions \citep{Huang2022, Jun2023, Meneses2024, Varadarajan2021, Zha2019}, and synthetic MRI generation \citep{Jacobs2024, Borges2023, Borges2025, Borges2026, Lpke2025}. Recently, self-supervised physics-guided models have been adapted to train directly on conventional MRI to infer quantitative maps \citep{Moya-Sez2021ISMRM, Qiu2024, Qiu2024ISMRM, vanLune2025}, eliminating the need for paired reference maps during training. However, these models have primarily been studied on small, clinically homogeneous conventional MRI datasets — typically comprising tens to a few hundred subjects. Moreover, the conventional MRIs in these studies were generally acquired using fixed protocols. While these exploratory studies represent an essential first step, it remains unclear whether these methods are robust and scalable to the diversity of real-world clinical imaging, characterized by variability in scanner hardware and sequence parameters.

In this work, we investigate the application of a self-supervised physics-guided deep learning framework to a large-scale, clinically heterogeneous dataset to infer quantitative T1, T2, and PD maps. We hypothesize that a model trained on heterogeneous conventional MRI data will learn robust, generalizable quantitative mappings across scanner systems, time, and sequence parameters. To this end, we assembled a dataset of over 4,000 clinical MRIs, containing T1w, T2w, and FLAIR sequences, acquired with a range of different sequence parameters. We trained the model on this dataset and evaluated its performance on a test set of over 600 scan sessions. Additionally, we assessed within-subject reproducibility of the generated quantitative maps across scanner hardware and sequence parameters. 

\section{Methods}
\label{methods}

\subsection{Dataset}
\label{methods:dataset}

\subsubsection{Large-scale clinical dataset}
We retrospectively collected clinical conventional MRI data from the radiology department at the University Medical Center Utrecht (UMCU), spanning the full range of neuro-imaging protocols. The collected data was acquired between 2018 and 2023. We selected MRIs obtained on $3$~T systems that featured T1w, T2w, and T2-FLAIR sequences, as these are the most commonly used contrasts in routine clinical practice. We did not include T1w acquisitions with gadolinium enhancement. T2w and T2-FLAIR scans were frequently acquired after gadolinium injection. We included these in the dataset because they are typically obtained immediately after injection, when gadolinium uptake is limited, and its impact on T2-weighted image contrast is minimal \cite{Vymazal2024}. All data were collected and processed in accordance with the regulations and ethical guidelines of our institution.

This data collection process yielded a large clinically heterogeneous dataset comprising 4,121 MRI sessions from 1,786 unique subjects (age range = 11–89 years; mean = 52.4 ($\pm$ 16.2 SD) years; 54.9\% male, 45.1\% female). The dataset included patients with a range of conditions, including but not limited to oncology, neurodegenerative, epilepsy, and vascular conditions. Scans were performed on four different Philips $3$~T systems: Achieva (34.1\%), Ingenia (27.5\%), Ingenia CX (21.8\%), and Ingenia Elition X (16.6\%). Sequence types were consistent across the dataset: 3D Spoiled Gradient-Echo (Spoiled GRE) for T1w, 2D Turbo Spin Echo (TSE) for T2w, and 3D TSE for T2-FLAIR. Sequence parameters were extracted from the DICOM headers and are summarized in Table \ref{tab:clinical_archive_info}.

\begin{table}[b]
\centering
\renewcommand{\arraystretch}{1.2}
\caption{Sequence parameters and information of the conventional brain MRIs in the assembled clinical dataset. For each contrast (T1w, T2w, and FLAIR) the presence of gadolinium, ranges of repetition time (TR), echo time (TE), inversion time (TI), flip angle (FA), bandwidth (BW), reconstruction matrix size (Rec.\ matrix) and acquisition duration (Acq.\ time) is listed. Spoiled GRE: Spoiled Gradient-Echo; TSE: Turbo Spin Echo; AP: Anterior--Posterior; RL: Right--Left; CC: Craniocaudal.}

\label{tab:clinical_archive_info}
\begin{tabular}{rccc}
\toprule
 & \textbf{T1w: 3D Spoiled GRE} & \textbf{T2w: 2D TSE} & \textbf{T2-FLAIR: 3D TSE} \\
\midrule
\textbf{Gadolinium} & No & Yes (98.1\%) & Yes (35.9\%) \\
\textbf{TR [ms]} & 5.03 -- 5.40 & 3848 -- 4557 & 4800 \\
\textbf{TE [ms]} & 2.247 -- 2.493 & 80 & 280 -- 370 \\
\textbf{TI [ms]} & NA & NA & 1650 \\
\textbf{FA [$^\circ$]} & 10 & 90 & 90 \\
\textbf{BW [Hz/px]} & 539 -- 543 & 184 -- 206 & 943 -- 1468 \\
\textbf{Rec.\ matrix (in-plane, AP $\times$ RL)} & 480 -- 560 $\times$ 480 -- 560 & 512 -- 560 $\times$ 512 -- 560 & 240 -- 288 $\times$ 240 -- 288 \\
\textbf{Rec.\ matrix (slices, CC)} & 306 -- 480 & 35 -- 42 & 240 -- 333 \\
\textbf{Acq.\ time [min:s]} & 1:59 -- 4:05 & 0:54 -- 1:29 & 3:17 -- 8:14 \\
\bottomrule
\end{tabular}
\end{table}

\subsubsection{Data preprocessing}
All scans were preprocessed using the same pipeline for consistency. First, we converted the scans from DICOM to NIfTI format using dcm2niix \citep{Li2016}, with sequence parameters extracted from the DICOM headers. Second, for each scan session, we rigidly registered the 3D T1w and 3D T2-FLAIR images to the 2D T2w image using ANTsPy \citep{Avants2009}. Third, all three conventional weighted images were resampled to an in-plane resolution of 1~$\times$~1~mm$^2$ while preserving the original slice spacing of the T2w acquisition ($4$~mm for all T2w acquisitions). Additionally, the axial slices of the images where center-cropped or zero-padded to a resolution of 224~$\times$~224. Fourth, we applied N4 bias-field correction \citep{Tustison2010} to all scans. Finally, superior and inferior slices with low signal-to-noise ratio were excluded to reduce artifacts. 

The clinical dataset was divided into training, validation, and test sets on a subject basis, with an 80/5/15\% split. This split resulted in 3,312 training sessions from 1,436 unique subjects, 206 validation sessions from 93 unique subjects, and 603 test sessions from 257 unique subjects. For tissue-specific evaluation of the test set, registered T1w images were skull-stripped with HD-BET \citep{Isensee2019} and segmented using FSL-FAST \citep{Zhang2001} into white matter (WM) and gray matter (GM). To minimize partial volume effects from cerebrospinal fluid (CSF), the WM and GM masks adjacent to CSF within $2.0$~mm were eroded.

\subsection{Model}
\label{methods:model}
We developed a physics-guided self-supervised 2D convolutional neural network (CNN) framework. The goal of this model is to generate quantitative T1, T2, and PD maps directly from conventional MRI without relying on paired quantitative reference data. The framework is made physics-guided by incorporating analytical closed-form MRI physical signal models into the learning process, building on prior work \citep{Moya-Sez2021ISMRM, Qiu2024, Qiu2024ISMRM, vanLune2025}.

\begin{figure}[tb]
    \centering
    \includegraphics[width=0.9\linewidth]{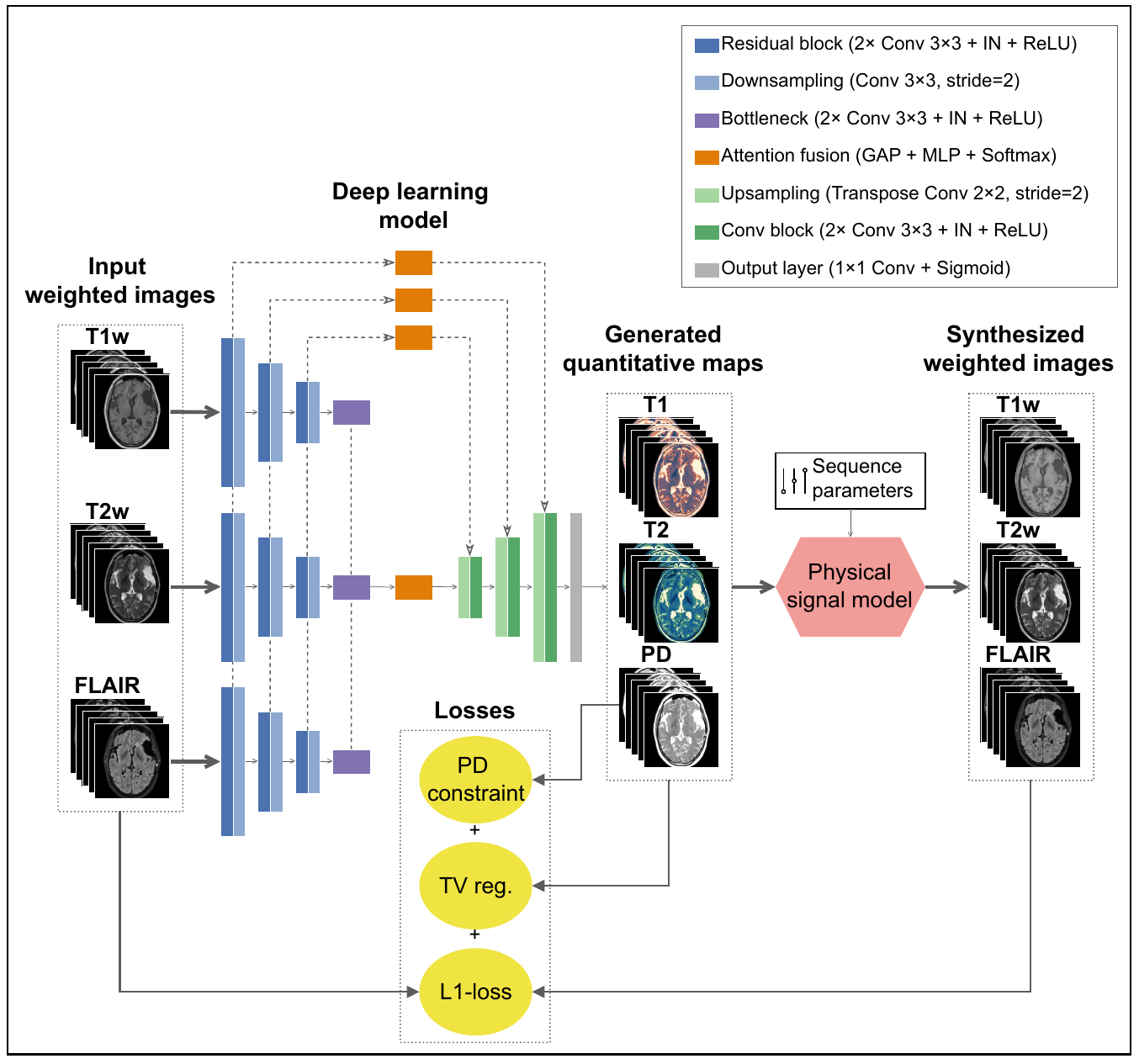}
    \caption{Overview of the proposed self-supervised physics-guided CNN framework. Each input contrast (T1w, T2w, FLAIR) is processed independently by a three-level (+~bottleneck level) shared-weight encoder. At each level, an attention fusion module applies global average pooling (GAP) and a shared multilayer perceptron (MLP). Subsequently, attention weights are computed using a across contrasts. The decoder outputs three channels (T1, T2, PD maps) from the fused features. From these quantitative maps, conventional MRI is synthesized using Bloch-based physical signal models. To train the deep learning model, an L1-loss is computed between the input and synthesized images. Additionally, total variation regularization (TV reg.) is applied to all quantitative maps, along with a soft lower bound on the PD map (PD constraint).}
    \label{fig:architecture}
\end{figure}

\subsubsection{Deep learning architecture}
The deep learning CNN follows a three-level, 2D U-Net–like design \citep{Ronneberger2015} that is capable of processing multi-contrast MRI. We chose CNNs over voxel-wise models to leverage spatial context and reduce sensitivity to noise \cite{Ulyanov_2020}. \added{A 2D model was chosen due to the anisotropic input data in the through-plane direction.} The model takes conventional MRIs as input (224~$\times$~224 axial brain slices) and outputs three channels (each channel 224~$\times$~224) representing the quantitative T1, T2, and PD maps (Figure \ref{fig:architecture}).

A shared-weight encoder processes each conventional MRI contrast independently. Each encoder level consists of a residual block with two 3~$\times$~3 convolutions, each followed by instance normalization (IN) and ReLU activation. Downsampling is performed via a strided 3~$\times$~3 convolution. The encoder feature widths are 64, 128, and 256 channels, respectively. The bottleneck doubles the feature dimensionality to 512 channels and applies two convolution–IN–ReLU blocks. At each level, feature maps from different input contrasts are combined through an attention-based fusion module, inspired by squeeze-and-excitation networks \cite{hu2019squeezeandexcitationnetworks}. First, spatial global average pooling (GAP) is applied to the feature maps of each contrast separately. Second, the resulting features are passed through a shared two-layer multilayer perceptron (MLP) to compute attention logits. These logits are then normalized with a softmax across contrasts to produce attention weights, which are used to weight the corresponding feature maps. Finally, the weighted features are summed to create a single fused representation at each U-Net level. This mechanism computes dynamic, contrast-specific attention weights that modulate the contribution of features of each input contrast at every U-Net level. Additionally, by using a shared-weight encoder and a softmax-normalized fusion module, the architecture is inherently flexible to the number of input contrasts; it can process any combination of input contrasts by redistributing attention weights among the available feature maps.

The decoder operates on the fused feature representations to generate the quantitative maps. In the decoder, the features are upsampled using transposed convolutions and concatenated with the corresponding fused skip connections from the encoder. Each decoder level applies convolution-IN–ReLU blocks to refine the upsampled features. Finally, a 1~$\times$~1 convolution maps the 64-channel decoder features to three output channels (T1, T2, PD), followed by a sigmoid activation. 
The sigmoid outputs are subsequently linearly scaled to the following parameter ranges: $[0, 1]$~a.u.\ for PD, $[0, 5]$~s for T1, and $[0, 3]$~s for T2. These ranges cover the full spectrum of in vivo values for human brain tissue \cite{Bojorquez2017, Stanisz2005}. In total, the network contains approximately 8.6 million trainable parameters.

\subsubsection{Physics-guided self-supervision}
The deep learning model is trained using a physics-guided self-supervised strategy. During training, the predicted quantitative maps are used to synthesize conventional weighted MRI via Bloch-based signal models \citep{Warntjes2008} parameterized by each scan’s acquisition settings (Figure~\ref{fig:architecture}). A self-supervised loss is computed between the synthesized and corresponding input images. The embedded physics model provides implicit supervision, removing the need for paired quantitative reference labels. The Bloch-based signal models are given in Supplementary Materials \ref{sec:appendix_signal_models}.

\paragraph{TSE Correction}
The analytical Bloch-based equations are generally less accurate for TSE acquisitions, because they use long echo trains with variable refocusing flip angles. This causes the effective T2-weighting to differ from that predicted by the nominal echo time (TE). To account for this, TE correction factors were calculated for the 2D T2w TSE and 3D T2-FLAIR TSE sequences, following the work of \citet{Busse2006}. The resulting factors (0.90 for T2w and 0.42 for FLAIR) were used to adjust the nominal TE, yielding a more realistic signal representation.

\paragraph{Input normalization and global scaling}
Conventional MRI acquisitions provide image intensities in arbitrary, non-standardized units. This results in signal intensities that are scaled by an unknown scan-dependent global factor. Resolving this scaling issue is necessary to align the acquired data with the physical signal model's output intensities. To address this, all conventional input images (T1w, T2w, and FLAIR) were normalized by dividing voxel intensities by the global histogram mode of the whole volume after preprocessing. We chose this normalization procedure because the histogram mode corresponds to WM intensities. This normalization procedure was coupled to a global intensity scaling procedure of the synthesized weighted images. Global scaling is necessary to compensate for differences in scaling between the normalized input weighted images and the synthesized weighted images when calculating the self-supervised loss. During training, for each subject and contrast, an optimal scaling factor $k$ was determined by solving a closed-form least-squares minimization problem between the input image ($I$) and the synthesized image ($\hat{I}$). The solution to this least-squares problem yields the fitted scaling factor $k$, where the index $i$ refers to an individual voxel within the image volume (Equation \ref{eq:scaling_factor_argmin}).

\begin{equation}
    \label{eq:scaling_factor_argmin}
    \underset{k}{\operatorname{arg\ min}} \left\lVert I - k\, \hat{I} \right\rVert^2  \Rightarrow
     k = \frac{\sum_i I_i \hat{I}_i}{\sum_i \hat{I}_i^2}
\end{equation}

To avoid arbitrary scaling we adopted a strategy similar to the work of \citet{Qiu2024}: the fitted scaling factor ($k$) was constrained to remain within $\pm~$20\% of an analytic reference. Because the input data was normalized to their global mode (representing WM intensity), we calculated the reference signal ($S_{WM}$) using the Bloch-based signal models with representative WM tissue parameter values (PD = $0.70$~a.u., T1 = $850$~ms, T2 = $70$~ms) and the specific sequence parameters parameters (TR,TE,TI,FA) for each subject and contrast. The reference scaling factor was then defined as the ratio between this $S_{WM}$ and the normalized input contrast mode.

\paragraph{Training losses and regularization}
The primary training objective was an L1-loss (mean absolute error) computed between the globally rescaled synthesized image ($\hat{I}$) and the normalized input image ($I$) for each contrast (T1w, T2w, and FLAIR). The L1-loss, defined in Equation \ref{eq:L1_loss}, was chosen because it mitigates the impact of large residuals arising from factors such as image artifacts, noise, imperfect registration, and imperfect global scaling.

\begin{equation}
\label{eq:L1_loss}
\mathcal{L}_{L_1}
=
\frac{1}{3}
\sum_{m \in \{\mathrm{T1w},\,\mathrm{T2w},\,\mathrm{FLAIR}\}}
\left\lVert I_m - \hat{I}_m \right\rVert_1 
\end{equation}

To suppress noise in the generated quantitative maps ($Q_M$), we applied isotropic total variation (TV) regularization. As shown in Equation \ref{eq:tv}, $TV(Q_M)$ is calculated by taking the L2-norm of the spatial gradient $\nabla Q_i$ at each voxel $i$. 

\begin{equation}
\label{eq:tv}
TV(Q_M) = \frac{1}{N} \sum_{i=1}^{N}\left\lVert \nabla Q_i \right\rVert_2, \qquad \mathcal{L}_{TV} = \frac{1}{3} \sum_{m \in \{\mathrm{PD},\,\mathrm{T1},\,\mathrm{T2}\}} TV(Q_m)
\end{equation}

The synthesis of quantitative maps from conventional MRI represents an ill-posed inverse problem, as multiple configurations of tissue parameters could theoretically result in similar qualitative intensities. To narrow the solution space of this ill-posed problem, a soft lower bound was imposed on the PD map (Equation \ref{eq:pd-constraint}). with the threshold $\tau$ set at 0.60 (similar to \citet{Qiu2024}).

\begin{equation}
\label{eq:pd-constraint}
\mathcal{L}_{\mathrm{PD\text{-}constraint}}
=
\frac{1}{N}
\sum_{i=1}^{N}
\max\!\left(0,\, \tau - Q_{\mathrm{PD},i}\right)
\end{equation}

The final training objective combined the three terms as shown in Equation \ref{eq:total_loss}:

\begin{equation}
\label{eq:total_loss}
\mathcal{L}
=
\mathcal{L}_{L_1}
+
\lambda_{TV}\,\mathcal{L}_{TV}
+
\lambda_{PD}\,\mathcal{L}_{\mathrm{PD\text{-}constraint}}
\end{equation}

For implicit regularization, a contrast dropout strategy was used, randomly omitting one input contrast (T1w, T2w, or FLAIR) from the deep learning model with a probability of 50\%. This approach prevents the model from over-relying on any single input contrast and encourages it to leverage complementary information across the different contrasts \cite{Borges2023}. This strategy is easily implemented within our framework due to the shared-weight encoder and attention-based fusion mechanism.

\subsubsection{Training procedure, hyperparameter tuning and implementation details}
The network was trained on the clinical dataset for 50 epochs using the Adam optimizer \citep{Kingma2014} with momentum parameters $\beta_1 = 0.9$ and $\beta_2 = 0.999$. To identify the optimal model configuration, hyperparameter optimization was performed using the Optuna framework \citep{Akiba2019}. The search space and explored values are summarized in Table~\ref{tab:optuna_hyperparams} in the Supplementary Materials. \replaced{Model selection was based on minimizing validation loss while ensuring quantitative outputs matched physiologically expected ranges for human brain tissue \cite{Bojorquez2017}.}{Model selection was based on stable convergence of the validation loss and stable convergence to physiologically plausible ranges of quantitative T1, T2, and PD values.} \added{Furthermore, the necessity of each regularization term and constraint was systematically tested through an ablation study, detailed in Supplementary Materials \ref{sec:appendix_ablation_study}.} The final model was trained using a learning rate of $1\times10^{-3}$, a batch size of 16 slices, and regularization weights $\lambda_{TV}=0.01$ and $\lambda_{PD}=0.1$. Network weights were initialized using Kaiming initialization \citep{He2015}. Each training batch consisted of slices randomly sampled from the entire training dataset. The final model weights were taken from the epoch with the lowest validation loss.

The implementation was developed in Python 3.11.9 using PyTorch 2.1.2 \citep{Paszke2019}, with data loading, preprocessing, and slice sampling handled by TorchIO 0.20.6 \citep{Prez-Garca2021}. Training was performed on a single NVIDIA Tesla V100 GPU and took approximately 30 hours.

\subsection{Evaluation}
Model performance was evaluated using a two-step strategy. First, we assessed the behavior of the model at scale on the clinical test set. Second, we performed a subject-specific reproducibility analysis of the generated quantitative maps.

\begin{table}[bt]
\centering
\renewcommand{\arraystretch}{1.2}
\caption{Characteristics of the included subjects from the clinical test set for subject-specific analysis. Listed are age, sex, whether the two sessions were acquired on the same or different scanner systems, the brain pathology, and any ongoing treatment.}
\label{tab:retrospective_scanners_pathology_treatment}
\begin{tabularx}{\linewidth}{l c c c X X}
\toprule
\textbf{Subject} & \textbf{Age} & \textbf{Sex} & \textbf{Scanner system} & \textbf{Pathology} & \textbf{Treatment} \\
\midrule

1 & 66 & F & 
Same &
Diffuse astrocytoma, IDH-mutant (WHO CNS5 grade 4) &
Chemotherapy (temozolomide) \\

2 & 77 & M &
Different &
Metastases from lung carcinomas &
None \\[4pt]

3 & 78 & M &
Different &
Metastases from melanoma &
Immunotherapy (nivolumab) \\ 

4 & 35 & M &
Different &
Anaplastic astrocytoma, IDH-mutant (WHO CNS5 grade 3) &
Chemotherapy (temozolomide) + anti-epileptics \\
\bottomrule
\end{tabularx}
\end{table}

\begin{table}[bt]
\centering
\renewcommand{\arraystretch}{1.2}
\caption{Sequence parameters for the two different protocols (P1 and P2) for the acquisitions of the healthy volunteer on a Philips Ingenia CX system. For each contrast (T1w, T2w, and FLAIR) the presence of gadolinium, repetition time (TR), echo time (TE), inversion time (TI), flip angle (FA), bandwidth (BW), reconstruction matrix size (Rec.\ matrix) and acquisition duration (Acq.\ time) is listed. Spoiled GRE: Spoiled Gradient-Echo; TSE: Turbo Spin Echo; AP: Anterior--Posterior; RL: Right--Left; CC: Craniocaudal.}
\label{tab:prospective_acquisitions_info}
\begin{tabular}{rcccccc}
\toprule
 & \multicolumn{2}{c}{\textbf{T1w: 3D Spoiled GRE}} 
 & \multicolumn{2}{c}{\textbf{T2w: 2D TSE}}
 & \multicolumn{2}{c}{\textbf{T2-FLAIR: 3D TSE}} \\
\cmidrule(lr){2-3}\cmidrule(lr){4-5}\cmidrule(lr){6-7}
 & P1 & P2
 & P1 & P2
 & P1 & P2 \\
\midrule
\textbf{Gadolinium} & No & No & No & No & No & No \\
\textbf{TR [ms]} & 5.33 & 5.10 & 3906 & 4990 & 4800 & 4800 \\
\textbf{TE [ms]} & 2.44 & 2.30 & 80 & 80 & 298 & 351 \\
\textbf{TI [ms]} & NA & NA & NA & NA & 1650 & 1650 \\
\textbf{FA [$^\circ$]} & 10 & 10 & 90 & 90 & 90 & 90 \\
\textbf{BW [Hz/px]} & 539 & 539 & 194 & 194 & 1276 & 801 \\
\textbf{Rec.\ matrix (AP $\times$ RL)} & 480 $\times$ 480  & 480 $\times$ 480  & 512 $\times$ 512 & 512 $\times$ 512 & 240 $\times$ 240 & 240 $\times$ 240 \\
\textbf{Rec.\ matrix (CC)} & 360 & 380 & 35 & 39 & 300 & 317 \\
\textbf{Acq.\ Time [min:s]} & 3{:}42 & 2{:}11 & 1{:}57 & 2{:}10 & 7{:}31 & 4{:}19 \\
\bottomrule
\end{tabular}
\end{table}

\subsubsection{Clinical archive test set}
We first evaluated the model on the clinical archive test set comprising 603 scan sessions of 257 unique subjects. For each session, quantitative T1, T2, and PD maps were generated from the input conventional MRI. For every scan session, we computed the mean quantitative value (for T1, T2 and PD) within the segmented WM and GM masks. Population-level metrics were subsequently derived by calculating the overall mean and standard deviation (SD) of these session-wise means. Additionally, distributions of T1, T2, and PD values were calculated for GM and WM across sessions. 

To assess the invariance of the quantitative maps to scanner hardware and sequences parameters, we group the clinical test set by scanner system and by protocol cluster. These protocol clusters were defined using agglomerative hierarchical clustering on the TR and TE values from the T1w, T2w, and FLAIR scans, resulting in nine distinct groups (see Supplementary Materials \ref{sec:appendix_protocol_clustering} for clustering methodology and centroids). For each group, we computed per-session mean values within the WM and GM. To quantify invariance, we calculated the inter-group coefficient of variation (CV) for each quantitative parameter (T1, T2 and PD) in WM and GM. The CV was calculated as the standard deviation of the group means (the average of per-session means within a group) divided by the global mean (average of the group means).

\begin{figure}[bt]
    \centering
    \includegraphics[width=0.95\linewidth]{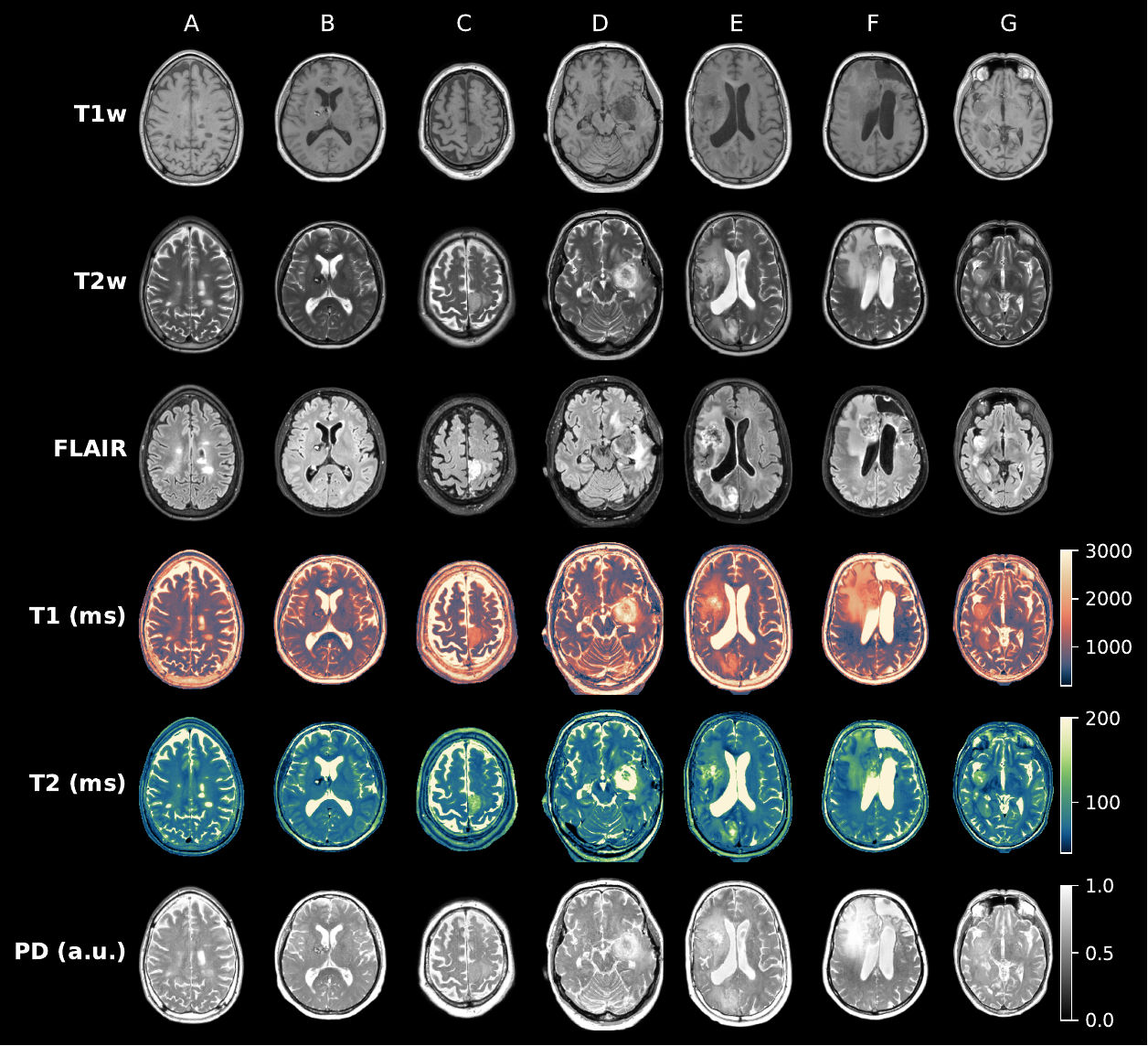}
    \caption{Representative example slices from the clinical archive test set. 
    For seven subject (A--G), featuring various lesions, the input contrasts (T1w, T2w, and FLAIR) are shown alongside the corresponding generated quantitative T1, T2, and PD maps.
    \textbf{(A)} Multiple sclerosis (MS); 
    \textbf{(B)} Cerebral cavernous venous malformation; 
    \textbf{(C)} Meningioma; 
    \textbf{(D)} Glioblastoma, status post-partial resection, chemoradiotherapy and post-radiation temozolomide; 
    \textbf{(E)} High-grade glioma progression, status post-radiotherapy and chemotherapy; 
    \textbf{(F)} Oligodendroglioma, status post-radiotherapy and chemotherapy; 
    \textbf{(G)} Multifocal high-grade glioma, status post-radiotherapy. 
    For patients A--C there was no ongoing treatment at the time of imaging.}
    \label{fig:test_set_slices}
\end{figure}

\subsubsection{Subject-specific reproducibility}
For the subject-specific analysis, we evaluated voxel-wise within-subject reproducibility across a subset of five subjects. Within-subject reproducibility assesses whether different scan sessions of the same subject yield similar quantitative values across scanner systems, protocols, and time intervals. This ensures that the estimated T1, T2, and PD reflect true tissue properties rather than non-physiological variation. The reproducibility analysis was performed in retrospective and prospective fashion. 

\paragraph{Retrospective reproducibility test}
Four subjects from the clinical test set were included for the reproducibility analysis. Only these four subjects met the predefined inclusion criteria: (1) two distinct scan sessions acquired as part of routine clinical follow-up; (2) an inter-scan interval $<3$ months; (3) no clinically relevant interval change in any lesion(s), as determined by the neuroradiologist in the team (JWD); and (4) adequate cross-session alignment to avoid excessive interpolation during registration for voxel-wise analysis. All included subjects had a brain lesion, and three received active treatment during the study period. A summary of subject characteristics is provided in Table \ref{tab:retrospective_scanners_pathology_treatment}, and sequence parameters for these scan sessions are listed in Table \ref{tab:mri_params_retrospective_subjects} in the Supplementary Materials. 

\paragraph{Prospective reproducibility test}
To complement these clinical data with a controlled assessment of sequence parameter sensitivity, prospective data were acquired from a healthy volunteer (male, 26 years) scanned on a $3$~T Philips Ingenia CX system. Written informed consent was obtained from the volunteer in accordance with the local institutional review board.  The subject was scanned twice using two different clinical protocols representative of the range of TR and TE values encountered in the clinical training dataset. These protocols were selected to sample the lower (approximately 20th percentile) and upper (approximately 80th percentile) portions of the clinical dataset parameter distribution (Table~\ref{tab:prospective_acquisitions_info}). These two protocols (P1 and P2) were acquired consecutively within the same scanning session. Prospective data were preprocessed using the same pipeline as the clinical data.

\paragraph{Voxel-wise analysis and reproducibility metrics}
For all five subjects, the two scan sessions were rigidly registered to the T2w image of the first session using ANTsPy \citep{Avants2009}. Quantitative T1, T2, and PD maps were generated from the registered conventional images. Reproducibility between sessions was assessed within the brain parenchyma (WM + GM) using three voxel-wise reproducibility metrics: Pearson correlation coefficient ($r$), concordance correlation coefficient ($CCC$), and mean voxel-wise relative difference.

\section{Results}

\subsection{Clinical archive test set}
Figure~\ref{fig:test_set_slices} shows representative example slices from the clinical archive test set. For seven subjects with diverse clinical pathologies — including neurodegenerative (A), vascular (B), and various oncology conditions (C--G) — the generated quantitative T1, T2, and PD maps are displayed alongside the corresponding input conventional MRIs (T1w, T2w, and FLAIR). The generated quantitative maps capture the lesion-related abnormalities visible on the conventional MRIs. Figure \ref{fig:app_synthetic_mri} in the Supplementary Materials illustrates conventional MRIs synthesized from the generated quantitative maps using the Bloch-based physical signal model for the same slices, alongside difference maps between the input and synthesized images.

\begin{figure}[tb]
    \centering
    \includegraphics[width=0.9\linewidth]{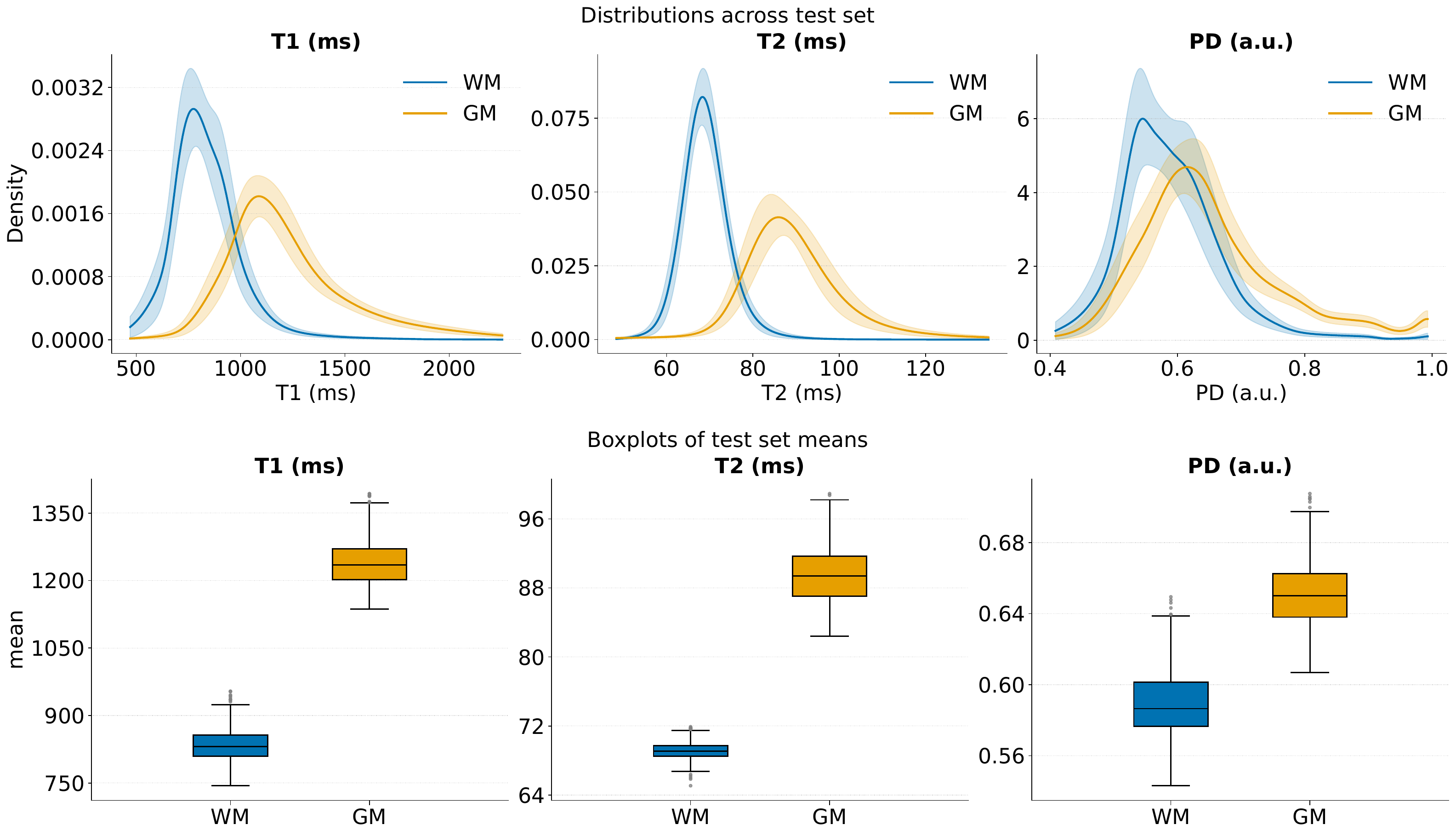}
    \caption{Summary of quantitative T1, T2, and PD values across all 603 clinical test sessions. Top: distributions of T1, T2, and PD values values in white matter (WM) and gray matter (GM). Solid lines indicate the mean value across sessions and shaded areas indicate one standard deviation. Bottom: boxplots of session-wise mean WM and GM values for T1, T2, and PD.}
    \label{fig:distribution_boxplot}
\end{figure}

Figure \ref{fig:distribution_boxplot} summarizes the quantitative values across all 603 test sessions. The generated maps show stable and well-separated WM and GM distributions. Across the full test set, the overall mean~$\pm$~SD values are: T1 (ms): WM = 834~$\pm$~37, GM = 1239~$\pm$~50; T2 (ms): WM = 69.1~$\pm$~1.0, GM = 89.5~$\pm$~3.3; and PD (a.u.): WM = 0.59~$\pm$~0.02, GM = 0.65~$\pm$~0.02. 

\begin{figure}[tb]
    \centering
    \includegraphics[width=0.9\linewidth]{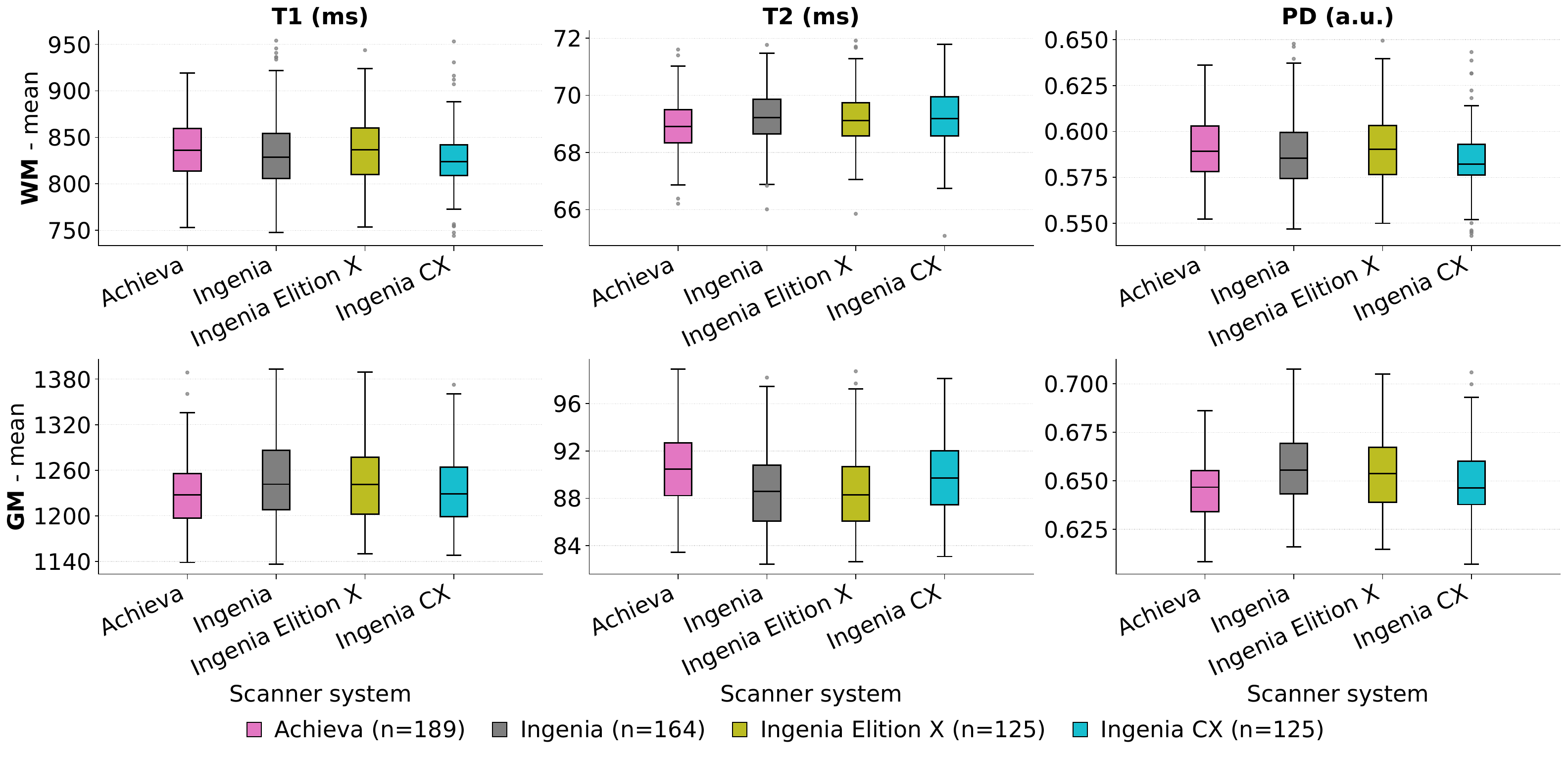}
    \caption{Boxplots of session-wise mean quantitative T1, T2, and PD values in WM and GM of the test set grouped by scanner system (Achieva, Ingenia, Ingenia CX, and Ingenia Elition X). The number of scan sessions in each group is displayed in the legend.}
    \label{fig:boxplot_scannermodel}
\end{figure}

\begin{figure}[tb]
    \centering
    \includegraphics[width=0.9\linewidth]{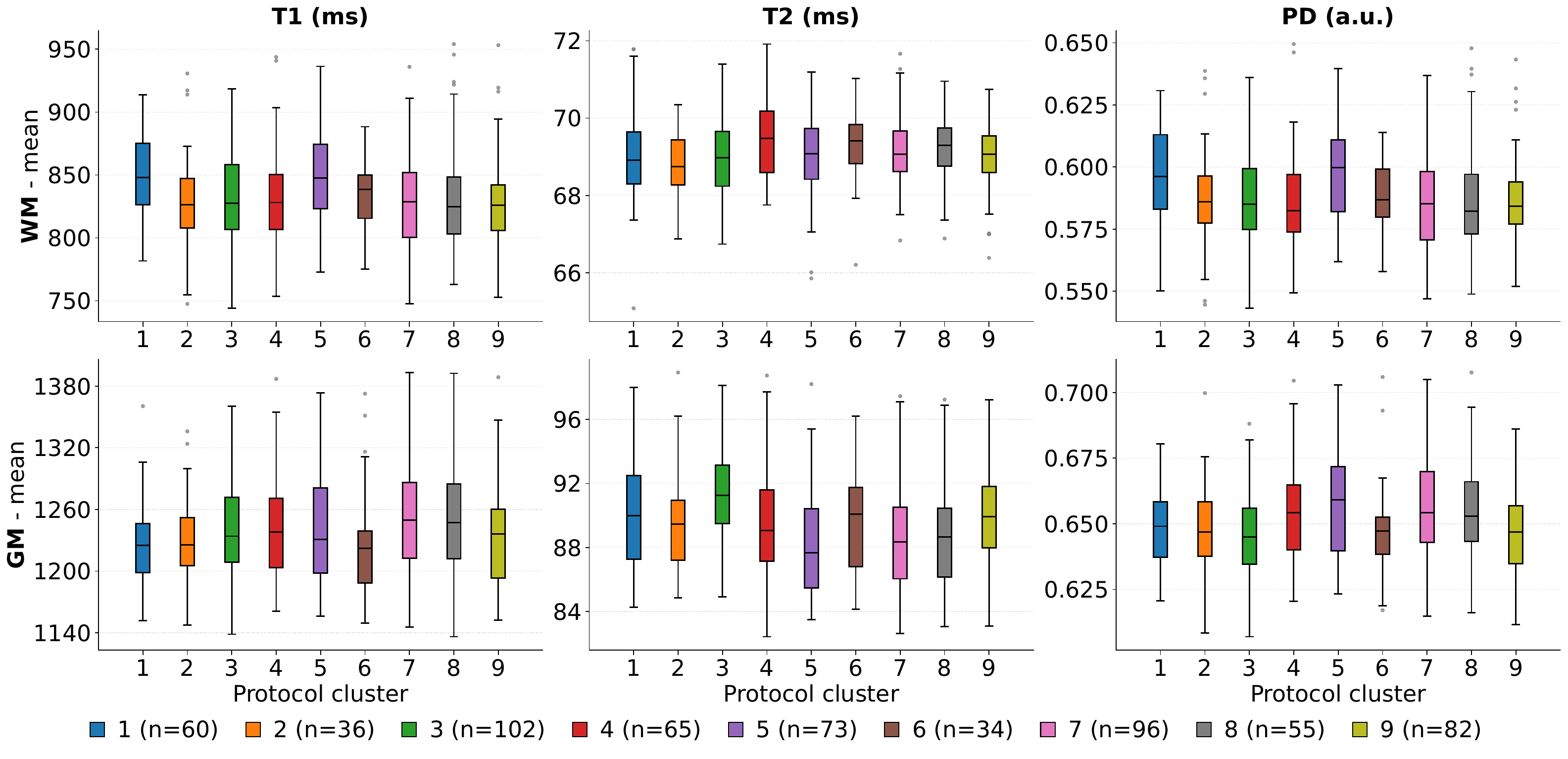}
    \caption{Boxplots of session-wise mean quantitative T1, T2, and PD values in WM and GM of the test set grouped by protocol cluster. The number of scan sessions in each group is displayed in the legend.}
    \label{fig:boxplot_protocolcluster}
\end{figure}

Boxplots grouped by scanner system and protocol cluster (Figures \ref{fig:boxplot_scannermodel} and \ref{fig:boxplot_protocolcluster}) show similar session-wise mean T1, T2, and PD values in WM and GM across these groups. Across scanner systems, the inter-group CV was 0.57\% (WM) and 0.68\% (GM) for T1, 0.22\% (WM) and 1.07\% (GM) for T2, and 0.54\% (WM) and 0.78\% (GM) for PD. For the nine protocol clusters, the inter-group CV was 1.08\% (WM) and 0.82\% (GM) for T1, 0.31\% (WM) and 1.07\% (GM) for T2, and 0.84\% (WM) and 0.72\% (GM) for PD.

\subsection{Subject-specific reproducibility}
We evaluated within-subject reproducibility using the five selected subjects.

\begin{figure}[tb]
    \centering
    \includegraphics[width=0.9\linewidth]{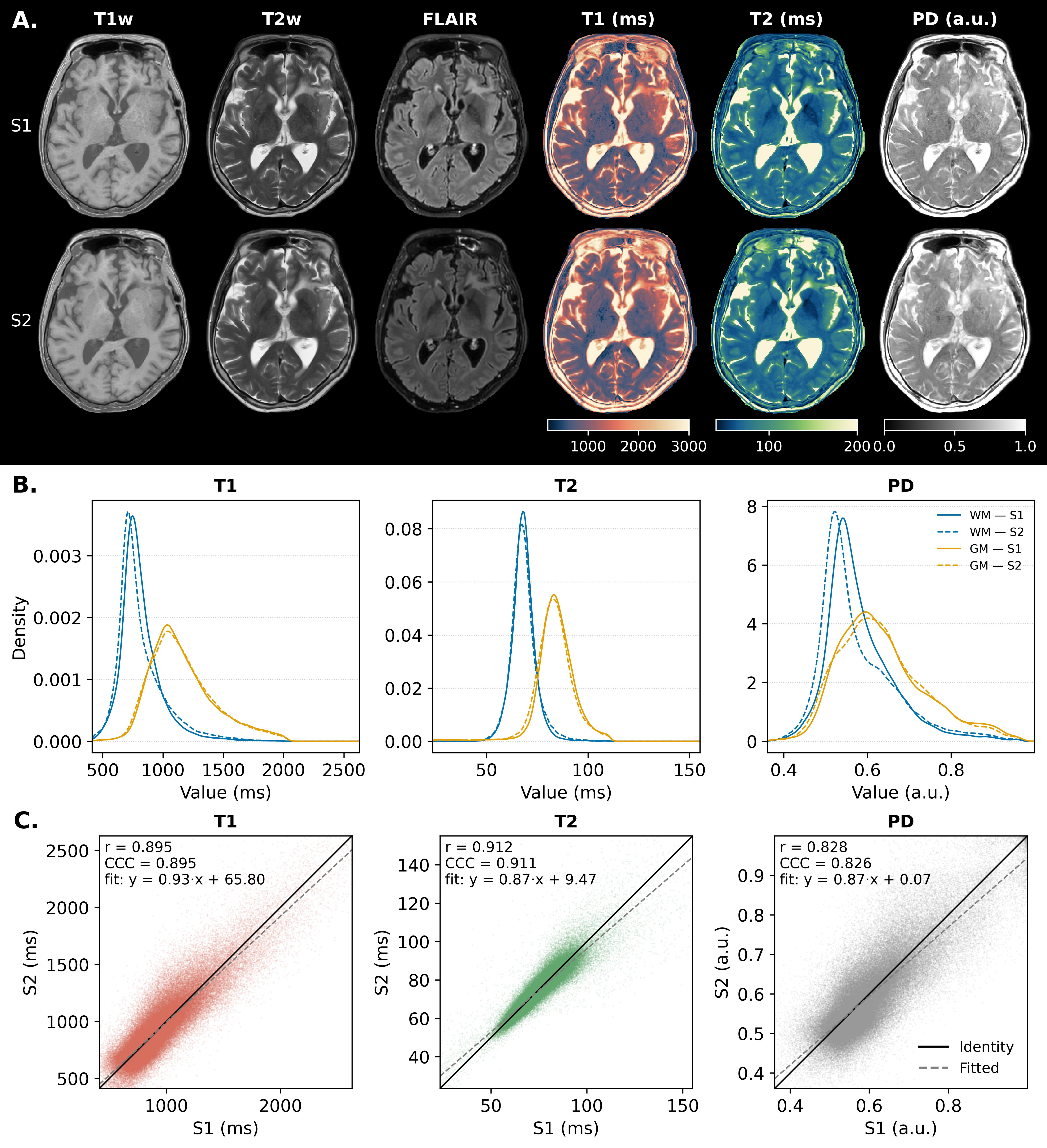}
    \caption{Within-subject reproducibility results for subject 1. 
            (A) Generated quantitative PD, T1, and T2 maps from the two scan sessions (S1 and S2), together with the corresponding input conventional images. The lesion is visible in the frontal region of the brain.
            (B) Whole-brain WM and GM value distributions for PD, T1, and T2. 
            (C) Voxel-wise scatterplots comparing S1 and S2 values for PD, T1, and T2 within brain parenchyma (WM + GM).}
    \label{fig:within_subject1}
\end{figure}

\begin{figure}[tb]
    \centering
    \includegraphics[width=0.9\linewidth]{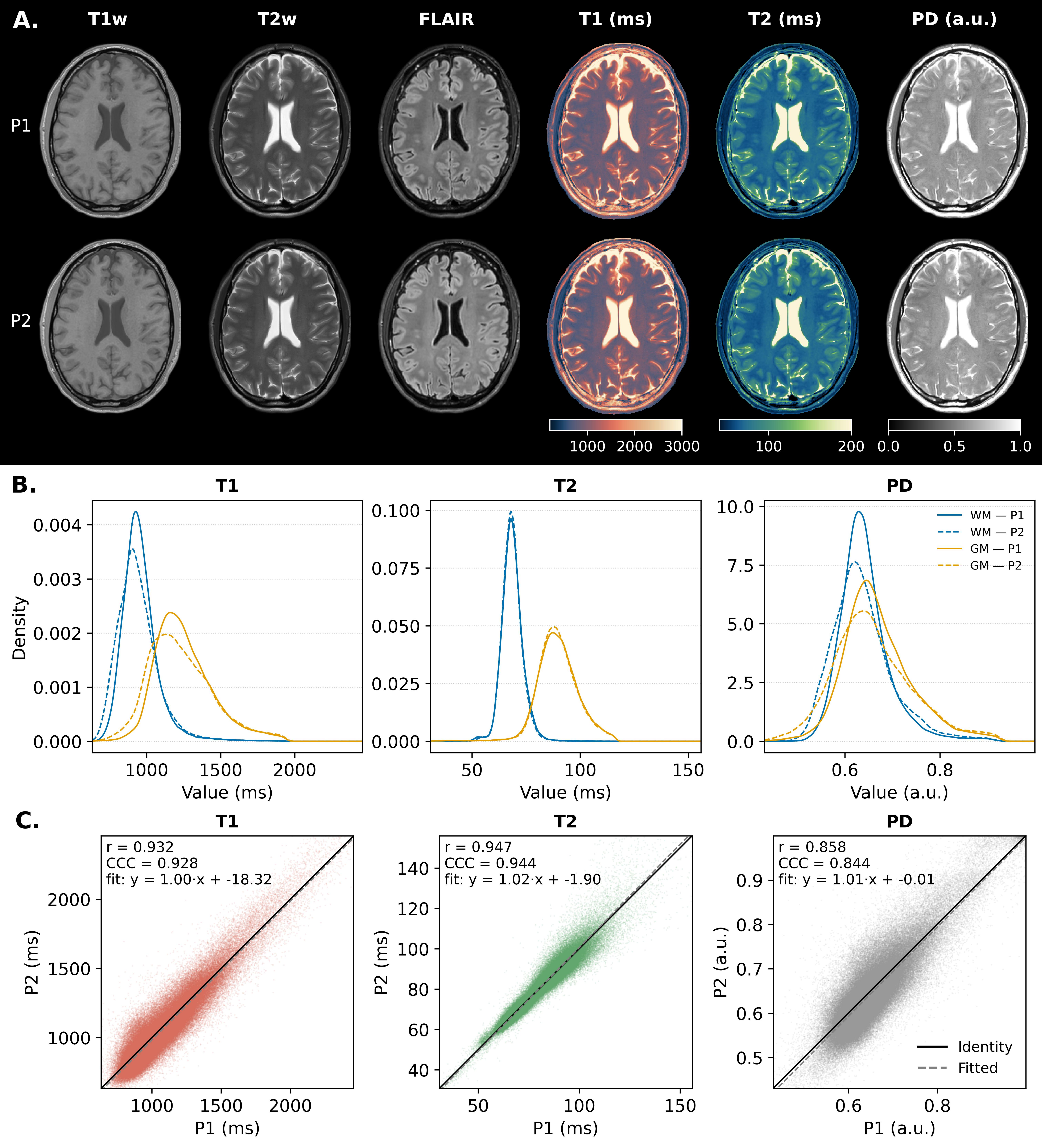}
        \caption{Prospective within-subject reproducibility in a healthy volunteer scanned with two different clinical protocols.
            (A) Generated quantitative PD, T1, and T2 maps from the two different protocols (P1 and P2), together with the corresponding input conventional images.
            (B) Whole-brain WM and GM value distributions for PD, T1, and T2. 
            (C) Voxel-wise scatterplots comparing S1 and S2 values for PD, T1, and T2 within brain parenchyma (WM + GM).}
    \label{fig:prospective_results}
\end{figure}

\paragraph{Retrospective reproducibility} 
Figure \ref{fig:within_subject1} illustrates results for Subject 1, included from the clinical test set. The generated quantitative T1, T2, and PD maps appeared visually consistent across sessions (Figure~\ref{fig:within_subject1}A). The whole-brain T1, T2 and PD distributions for WM and GM largely overlapped (Figure~\ref{fig:within_subject1}B). Voxel-wise scatterplots (Figure~\ref{fig:within_subject1}C) showed tight clustering along the identity line, with  Pearson $r$ and $CCC$ exceeding 0.89 for T1 and T2. Reproducibility figures for the remaining included subjects (Subjects 2-4) are provided in the Supplementary Materials (Figures \ref{fig:subject2}–\ref{fig:subject4}).

\paragraph{Prospective reproducibility}
Figure \ref{fig:prospective_results} illustrates results from the prospective volunteer data, scanned using two different protocols. The quantitative maps and whole-brain value distributions are presented in Figure \ref{fig:prospective_results}A and Figure \ref{fig:prospective_results}B, respectively. Voxel-wise scatterplots (Figure~\ref{fig:prospective_results}C) showed tight clustering along the identity line, with Pearson $r$ and $CCC$ exceeding 0.92 for T1 and T2, and exceeding 0.84 for PD. 

Voxel-wise reproducibility metrics for all five subjects are summarized in Table~\ref{tab:within_subject_repeatability}. Note that subjects 2–4 were scanned on different scanner systems for their two sessions. For all five subjects, Pearson $r$ and $CCC$ values exceeded 0.82 for T1 and T2. Mean relative voxel-wise differences were low across all quantitative parameters and subjects, especially for T2 (<~6\%). The prospective volunteer, where the two protocols were acquired consecutively, exhibited the lowest mean relative differences. Mean WM and GM quantitative values for all subjects are reported in the Supplementary Materials in Table \ref{tab:within_subject_means_compact_grid}. The largest observed difference in means was found for T1 WM in subject 4 (5.5\%), while T2 WM differences remained below 2\% across all subjects. The difference in means was especially low for the prospective volunteer study across all quantitative parameters (<~2\%).

\begin{table}[bt]
\centering
\caption{Voxel-wise within-subject reproducibility metrics for the five selected subjects (four included from the clinical test set and the healthy prospective volunteer). Pearson correlation coefficient ($r$), concordance correlation coefficient ($CCC$), and mean voxel-wise relative difference are reported for T1, T2, and PD within brain parenchyma (WM + GM).}
\label{tab:within_subject_repeatability}   
\renewcommand{\arraystretch}{1.2}
\setlength{\tabcolsep}{5pt}
\begin{tabular}{lccccccccc}
\toprule
\multirow{2}{*}{\textbf{Subject}} &
\multicolumn{3}{c}{\textbf{T1}} &
\multicolumn{3}{c}{\textbf{T2}} &
\multicolumn{3}{c}{\textbf{PD}} \\
\cmidrule(lr){2-4} \cmidrule(lr){5-7} \cmidrule(lr){8-10}
 & $r$ & $CCC$ & Mean diff. & $r$ & $CCC$ & Mean diff. & $r$ & $CCC$ & Mean diff. \\
\midrule
1   & 0.895 & 0.895 & 11.3\% & 0.912 & 0.911 & 3.9\% & 0.828 & 0.826 & 7.8\% \\
2  & 0.858 & 0.854 & 13.6\% & 0.895 & 0.859 & 5.6\% & 0.756 & 0.746 & 9.6\% \\
3  & 0.903 & 0.902 & 10.3\% & 0.921 & 0.920 & 4.1\% & 0.806 & 0.806 & 7.2\% \\
4  & 0.851 & 0.843 & 14.8\% & 0.839 & 0.824 & 5.1\% & 0.716 & 0.701 & 10.2\% \\
\midrule
Volunteer 
  & 0.932 & 0.928 & 6.9\% 
  & 0.947 & 0.944 & 2.9\% 
  & 0.858 & 0.844 & 5.1\% \\
\bottomrule
\end{tabular}
\end{table}

\section{Discussion}
In this work, we presented a physics-guided self-supervised deep learning framework that generates quantitative T1, T2, and PD maps directly from conventional clinical MRI. While earlier self-supervised frameworks for quantitative mapping from conventional MRI were limited to relatively small, clinically homogeneous research datasets with fixed sequence settings \citep{Moya-Sez2021ISMRM, Qiu2024, Qiu2024ISMRM, vanLune2025}, we demonstrate that such methods can scale effectively to large-scale, heterogeneous clinical data. Our framework was trained and evaluated on a broad dataset (>~4,000 scan sessions) covering all neuro-MRI protocols acquired at our institution over six years, spanning multiple MRI scanner systems and a range of sequence parameters. 

Across more than 600 clinical test sessions, the generated quantitative maps exhibited WM and GM value distributions whose means and standard deviations were consistent with literature reports \cite{Bojorquez2017, Stanisz2005, Choi2022}. When scan sessions were grouped by scanner system or by sequence parameter settings, similar WM and GM mean values were observed for T1, T2, and PD. Inter-group CV values were approximately 1\% or lower for all quantitative parameters in both groupings. These CV values are lower than those reported in repeatability studies of dedicated qMRI sequences acquired repeatedly in the same subject \cite{Buonincontri2019, Liu2025, Nunez-Gonzalez2021}. Thus, these results suggest that the generated quantitative maps from our proposed framework are invariant to the scanner hardware and sequence protocol changes present in our conventional MRI dataset.

Subject-specific analyses demonstrated excellent voxel-wise reproducibility across different scanner systems and sequence parameters. For the included subjects from the clinical test set, despite inter-scan intervals of approximately three months and the use of different scanner systems, we observed high voxel-wise correlation and agreement with Pearson $r$ and $CCC$ values exceeding 0.82 for both T1 and T2. Mean voxel-wise differences were low, particularly for T2 (approximately 5\%). To isolate the influence of sequence parameters from scanner system and biological variability, a prospective experiment was conducted where two distinct protocols were acquired consecutively. This controlled setup yielded even higher correlation and agreement ($r$,$CCC$~>~0.92 for T1/T2; ~>~0.82 for PD). Correspondingly, mean voxel-wise differences were also minimal. Together, these reproducibility findings indicate that the model effectively disentangles intrinsic relaxation properties from extrinsic scanner- and protocol-related variations.

Our findings are generally in line with reported inter-scanner reproducibility ranges for dedicated qMRI techniques. For example, \citet{Gracien2020} observed voxel-wise inter-scanner differences of up to 5.2\% for T1 and T2 over inter-scan intervals shorter than five weeks. \citet{Buonincontri2019} found multi-site reproducibility with CVs of 3–8\% for T1, 8–14\% for T2, and approximately 5\% for PD, together with high agreement ($CCC$ = 0.92–0.96) despite inter-scan intervals of 2–6 months. Interestingly, our framework demonstrated superior reproducibility for T2 compared to T1, reversing the trend reported in the literature \cite{Gracien2020, Buonincontri2019, Krzdrfer2019}. This likely reflects complementary information used as input, as the model uses two T2-weighted inputs (T2w and FLAIR), but a single T1-weighted input.

Earlier self-supervised physics-guided deep learning frameworks for quantitative mapping operated on skull-stripped inputs \citep{Moya-Sez2021ISMRM} and, in some cases, also relied on tissue-segmentation-based regularization during training \citep{Qiu2024, Qiu2024ISMRM}. In contrast, our framework requires only minimal preprocessing and is trained directly on non-skull-stripped, unsegmented clinical images. We also introduce a feature fusion mechanism designed to enable dynamic weighting of input-contrast features during quantitative map generation.

\added{To evaluate the specific regularization terms and constraints employed during training, an ablation study (Supplementary Materials \ref{sec:appendix_ablation_study}) assessed the necessity of the global scaling constraint, PD constraint, TV regularization, and contrast dropout during training. Omitting these components either skewed quantitative values outside established literature ranges (Supplementary Figures \ref{fig:ablation_combined_stacked} and \ref{fig:ablation_grid}) or reduced voxel-wise reproducibility (Supplementary Table \ref{tab:ablation_study}). Furthermore, exploratory results highlighted a promising future direction for the contrast dropout strategy: it not only improves model robustness against missing input contrasts during inference (Supplementary Figure \ref{fig:missing_modality}), but it could also facilitate the synthesis of those missing conventional MR images directly from the inferred quantitative maps.}

Some reflections on the validation strategy adopted in this study follow. Earlier studies — both self-supervised and supervised \citep{Qiu2024, Qiu2024ISMRM, Qiu2022, Moya-Sez2021, Sun2023} — typically validated inferred quantitative maps against a reference qMRI acquisition. However, measured relaxation times vary widely across qMRI techniques \citep{Bojorquez2017}, and no consensus qMRI acquisition protocol has been established \citep{Keenan2019, Asslnder2025, Saltarelli2025}. As a result, direct comparison with a reference qMRI method may obscure the model's true performance. In contrast, our evaluation emphasizes the invariance of the generated quantitative maps to scanner hardware and sequence parameters. Other studies have used datasets of comparable size for self-supervised physics-guided deep learning in MRI \citep{Borges2023, Borges2025, Borges2026}. In those works, the generated quantitative maps served primarily as intermediate representations for synthesizing conventional weighted images, rather than outputs of direct interest. In the present work, we directly analyze the quantitative parameter values themselves.

\subsection{Limitations}
\added{A primary limitation of this study is the scope of the dataset. Despite its large size and clinical heterogeneity, all data came from a single center. Additionally, the scans were acquired exclusively on $3$~T systems from a single vendor (Philips). Consequently, the model's ability to generalize to other sites, vendors, or magnetic field strengths is not yet established. This cross-site validation remains an important direction for future work.} Furthermore, this study focused on the most common neuro-MRI sequence types at our institution as model input (T1w 3D Spoiled GRE, T2w 2D TSE, and T2-FLAIR 3D TSE). However, other sequence types for these conventional scans are also used both within our institution and across centers. These sequence types may include, for example, spin echo and inversion-prepared gradient-echo sequences. Additionally, sequences can be implemented as 2D or 3D acquisitions. Supplementary Materials \ref{sec:appendix_multi_sequences} show that when multiple sequence types for the same conventional weighted images are combined during training, the resulting quantitative maps are not reproducible within subjects across input sequence types. This likely reflects the inherent limitations of the Bloch-based signal model, which does not account for factors such as B$_1$ inhomogeneity, magnetization transfer, partial-volume effects, or echo-train dynamics \cite{Saltarelli2025}. Addressing these limitations will require modeling the signal behavior of the different sequence types more accurately. Possible directions include adopting extended phase-graph formulations (e.g., Torch EPG-X \citep{CenciniM2024}), incorporating explicit B$_1^+$ and MT correction terms, or learning a more accurate signal model in a data-driven manner. Such extensions could enable harmonization of multi-center data acquired with diverse sequence types and scanner vendors, broadening the applicability of the proposed framework.

A common challenge in processing clinical MRI is the variation in acquisition geometry across different images. In clinical practice, 2D multi-slice acquisitions (e.g., T2w) are often combined with 3D acquisitions (e.g., T1w, FLAIR), each with differing resolutions. To address this, we registered all inputs to the 2D T2w geometry and resampled to a $1\times1$ mm in-plane resolution. This, however, reduced the through-plane resolution and caused loss of volumetric detail from the original 3D scans. Slice-thickness mismatches may also introduce partial-volume effects, which could complicate model deployment in multi-center settings with heterogeneous acquisition protocols. Another source of variability in clinical MRI is the presence of gadolinium contrast in certain scans. In our dataset, we retained T2w and T2-FLAIR scans acquired immediately after gadolinium injection. Because gadolinium primarily affects T1 relaxation and has minimal effect on T2 \cite{Vymazal2024}, its impact on these scans is expected to be limited. While residual bias in the generated quantitative maps cannot be entirely excluded, including these scans allowed us to maximize the use of acquired clinical data.

A current challenge in the field of self-supervised physics-guided quantitative mapping from conventional MRI is the global intensity scaling between the input and synthesized weighted images. Previous approaches either guided and constrained the scaling using acquired reference qMRI maps \citep{Qiu2024} or learned it in a weakly supervised manner from limited paired data \citep{Xu2025}. Such approaches require qMRI reference acquisitions and may generalize poorly beyond the small paired-data distribution. In our framework, we computed the optimal global scaling factor by solving a least-squares problem during training. Following \citet{Qiu2024}, we constrained the scaling within a predefined range (based on literature quantitative values) to prevent arbitrary scaling in this ill-posed problem. This approach avoids the need for any acquired reference qMRI data, while ensuring stable and meaningful loss computation during training. An alternative strategy would be to use scale-invariant losses (e.g., correlation-based) combined with physiological priors that regularize expected WM, GM, and CSF values. By avoiding a dependency on segmentation-based priors, our framework remains more robust to the high variability inherent in real-world clinical data.

Several limitations of this study's design should be acknowledged. First of all, in general, the analyses relied on tissue segmentations derived from clinical scans, which may be imperfect in the presence of lesions or other abnormalities. Moreover, clinical acquisitions may contain artifacts (e.g., motion) that affect both the input conventional scans and the generated quantitative maps. The within-subject reproducibility analyses also have specific limitations. All selected clinical subjects had underlying brain pathology, and three were undergoing active treatment, potentially introducing biological variability across sessions. Furthermore, voxel-wise comparisons are sensitive to registration inaccuracies and partial-volume effects. Finally, although the model was evaluated on a large clinical test set, the number of subjects available for subject-specific reproducibility analysis was limited; larger, dedicated datasets will be required to more comprehensively assess reproducibility.

\subsection{Outlook}

The results of this work indicate that the proposed approach robustly transforms conventional MRI into quantitative maps at scale. Our framework can be applied to the extensive clinical brain MRI archives available in most hospitals. Because the framework relies solely on routinely acquired T1w, T2w, and FLAIR scans, no additional scan time is required to obtain quantitative maps. Making these maps available at scale opens the door to large-scale investigations of quantitative biomarkers across diverse patient populations. These investigations may include applications such as diagnosis and longitudinal monitoring in brain tumors \citep{Ding2021, MohamedSajer2025, Chekhonin2024} and neurodegenerative diseases \cite{Lou2021}. Such investigations may, in turn, support and accelerate the clinical adoption of dedicated qMRI techniques. \added{As an initial indicator of this clinical potential, the generated maps successfully capture physiological changes within the diverse lesions shown in Figure~\ref{fig:test_set_slices}. For instance, compared to normal-appearing white matter, we observed T1 \& T2 prolongations of, respectively, 88.5\% \& 121.3\% in an MS lesion (Subject A), 83.1\% \& 89.8\% in a meningioma (Subject C), and 83.5\% \& 26.4\% in an oligodendroglioma (Subject F). These relative increases are similar to those reported in qMRI literature for these specific pathologies \cite{Blystad94, Schilder2025, Mostardeiro2021, Kikuchi2022}.} Furthermore, by transforming conventional MRI into quantitative maps, the framework effectively harmonizes data across different scanner hardware and sequence parameter settings. Such harmonization could improve the generalizability of downstream deep learning models for image analysis, which often struggle with variability in scanner hardware and acquisition parameters \cite{Bento2022, Abbasi2024}. Finally, although we demonstrated the framework on brain data, the framework could be extended to other anatomical regions, such as the prostate \cite{Schieda2021} and knee \cite{Atkinson2019}.


\section{Conclusion}
We presented a physics-guided self-supervised framework that robustly transforms conventional clinical MRI into reproducible quantitative T1, T2, and PD maps, without relying on dedicated qMRI reference data. By training on a large, heterogeneous dataset spanning six years, multiple scanners, and a range of sequence parameters, the framework generates quantitative maps robust to scanner- and protocol-related variability.

\section*{Acknowledgments}
This research was supported by the Hanarth Fonds for AI in Oncology. 


\bibliographystyle{unsrtnat}
\bibliography{references}  

\clearpage

\setcounter{page}{1}
\pagenumbering{roman} 

\title{Supplementary Materials of: Quantitative mapping from conventional MRI using self-supervised physics-guided deep learning: applications to a large-scale, clinically heterogeneous dataset}

\maketitle

\pagenumbering{arabic}
\setcounter{page}{1}

\appendix

\counterwithin{figure}{section}
\counterwithin{table}{section}
\counterwithin{equation}{section}

\section{Bloch-based signal models}
\label{sec:appendix_signal_models}

The Bloch-based signal models used for training the self-supervised physics-guided deep learning framework are given by the following equations (\ref{eq:t1w}, \ref{eq:t2w}, and \ref{eq:flair}):

\begin{equation}
    S_{\text{T1w Spoiled GRE}} = PD \cdot \sin(FA) \cdot \frac{1 - \exp(-TR/T1)}{1 - \cos(FA) \cdot \exp(-TR/T1)} \cdot \exp(-TE/T2)
    \label{eq:t1w}
\end{equation}

\begin{equation}
    S_{\text{T2w TSE}} = PD \cdot (1 - \exp(-TR/T1)) \cdot \exp(-TE/T2)
    \label{eq:t2w}
\end{equation}

\begin{equation}
    S_{\text{FLAIR TSE}} = |PD \cdot (1 - 2 \exp(-TI/T1) + \exp(-TR/T1)) \cdot \exp(-TE/T2)|
    \label{eq:flair}
\end{equation}

Here, PD denotes the proton-density, T1 and T2 are the longitudinal and transverse relaxation times, respectively. The sequence parameters are given by the flip angle (FA), repetition time (TR), echo time (TE), and inversion time (TI).

\section{Hyperparameter tuning of deep learning model}

Table \ref{tab:optuna_hyperparams} outlines the hyperparameter search space and the specific values explored during the optimization process to identify the optimal model configuration.

\begin{table}[h]
\centering
\caption{Hyperparameter search space and range of values evaluated during the optimization process.}
\label{tab:optuna_hyperparams}
\renewcommand{\arraystretch}{1.3} 
\begin{tabular}{ll} 
\hline 
\textbf{Hyperparameter} & \textbf{Explored values} \\
\hline
Learning rate & [0.00001, 0.0001, 0.001, 0.01]\\
$\lambda_{PD}$ & [0, 0.05, 0.1, 0.15, 0.20] \\
$\lambda_{TV}$ & [0, 0.0001, 0.001, 0.01, 0.1] \\ 
Weight initialization & [Xavier, Kaiming, Normal] \\
Batch size & [8, 16, 32] \\ 
\hline 
\end{tabular}
\end{table}

\section{Protocol clustering clinical archive test set}
\label{sec:appendix_protocol_clustering}

\begin{figure}[h]
    \centering
    \includegraphics[width=\linewidth]{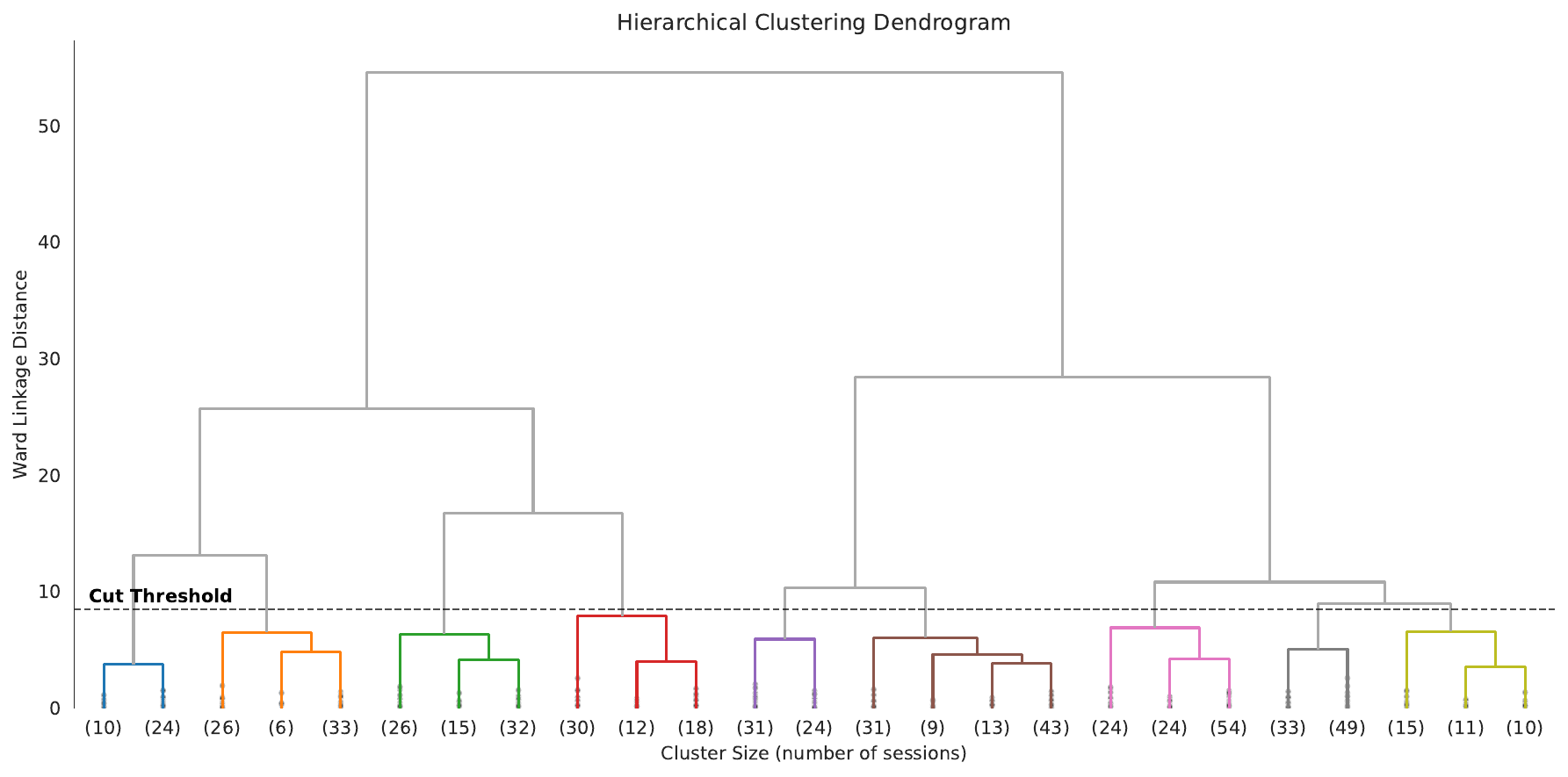}
    \caption{Hierarchical clustering dendrogram of sequence parameters. The dendrogram displays the hierarchical agglomerative clustering of 603 test sessions using z-score normalized repetition time (TR) and echo time (TE) for T1w, T2w, and FLAIR contrasts. The vertical axis represents the Ward linkage distance. A cut threshold (dashed horizontal line) was applied to define the 9 distinct protocol clusters. The values in parentheses on the x-axis represent the number of sessions (cluster size) within each leaf node.}
    \label{fig:app_dendrogram}
\end{figure}

To assess model robustness across diverse acquisition settings, we performed hierarchical agglomerative clustering (Ward’s linkage) on the z-score normalized TR and TE parameters of the T1w, T2w, and FLAIR scans of the test set (N=603). Visual inspection of the dendrogram (Fig. \ref{fig:app_dendrogram}) determined an optimal separation of 9 distinct protocol clusters. Table \ref{tab:app_cluster_means} summarizes the cluster centroids.

\begin{table}[h]
\centering
\caption{Centroids of sequence parameters for the nine protocol clusters identified in the clinical test set. The number of subjects ($N$) is listed for each cluster. Parameters include repetition time (TR), echo time (TE), inversion time (TI), and flip angle (FA).}
\label{tab:app_cluster_means}
\renewcommand{\arraystretch}{1.2}
\setlength{\tabcolsep}{2.5pt} 
\small
\begin{tabular}{@{}cc ccc ccc cccc@{}}
\toprule
\multirow{2}{*}{\textbf{Cluster}} & \multirow{2}{*}{$\boldsymbol{N}$} & \multicolumn{3}{c}{\textbf{T1w}} & \multicolumn{3}{c}{\textbf{T2w}} & \multicolumn{4}{c}{\textbf{FLAIR}} \\
\cmidrule(lr){3-5} \cmidrule(lr){6-8} \cmidrule(lr){9-12}
 & & \textbf{TR [ms]} & \textbf{TE [ms]} & \textbf{FA [$^\circ$]} & \textbf{TR [ms]} & \textbf{TE [ms]} & \textbf{FA [$^\circ$]} & \textbf{TR [ms]} & \textbf{TE [ms]} & \textbf{TI [ms]} & \textbf{FA [$^\circ$]} \\
\midrule
1 & 60 & 5.17 & 2.31 & 10 & 4315 & 80 & 90 & 4800 & 288 & 1650 & 90 \\
2 & 36 & 5.23 & 2.38 & 10 & 3955 & 80 & 90 & 4800 & 301 & 1650 & 90 \\
3 & 102 & 5.28 & 2.40 & 10 & 3853 & 80 & 90 & 4800 & 304 & 1650 & 90 \\
4 & 65 & 5.24 & 2.35 & 10 & 4313 & 80 & 90 & 4800 & 337 & 1650 & 90 \\
5 & 73 & 5.15 & 2.30 & 10 & 4340 & 80 & 90 & 4800 & 336 & 1650 & 90 \\
6 & 34 & 5.27 & 2.38 & 10 & 4331 & 80 & 90 & 4800 & 293 & 1650 & 90 \\
7 & 96 & 5.30 & 2.40 & 10 & 4143 & 80 & 90 & 4800 & 353 & 1650 & 90 \\
8 & 55 & 5.29 & 2.40 & 10 & 3905 & 80 & 90 & 4800 & 351 & 1650 & 90 \\
9 & 82 & 5.28 & 2.40 & 10 & 4062 & 80 & 90 & 4800 & 302 & 1650 & 90 \\
\bottomrule
\end{tabular}
\end{table}

\clearpage

\section{Synthesis of conventional MRI from generated quantitative maps}
\label{sec:appendix_synthetic_mri}

\begin{figure}[h]
    \centering
    \includegraphics[width=0.85\linewidth]{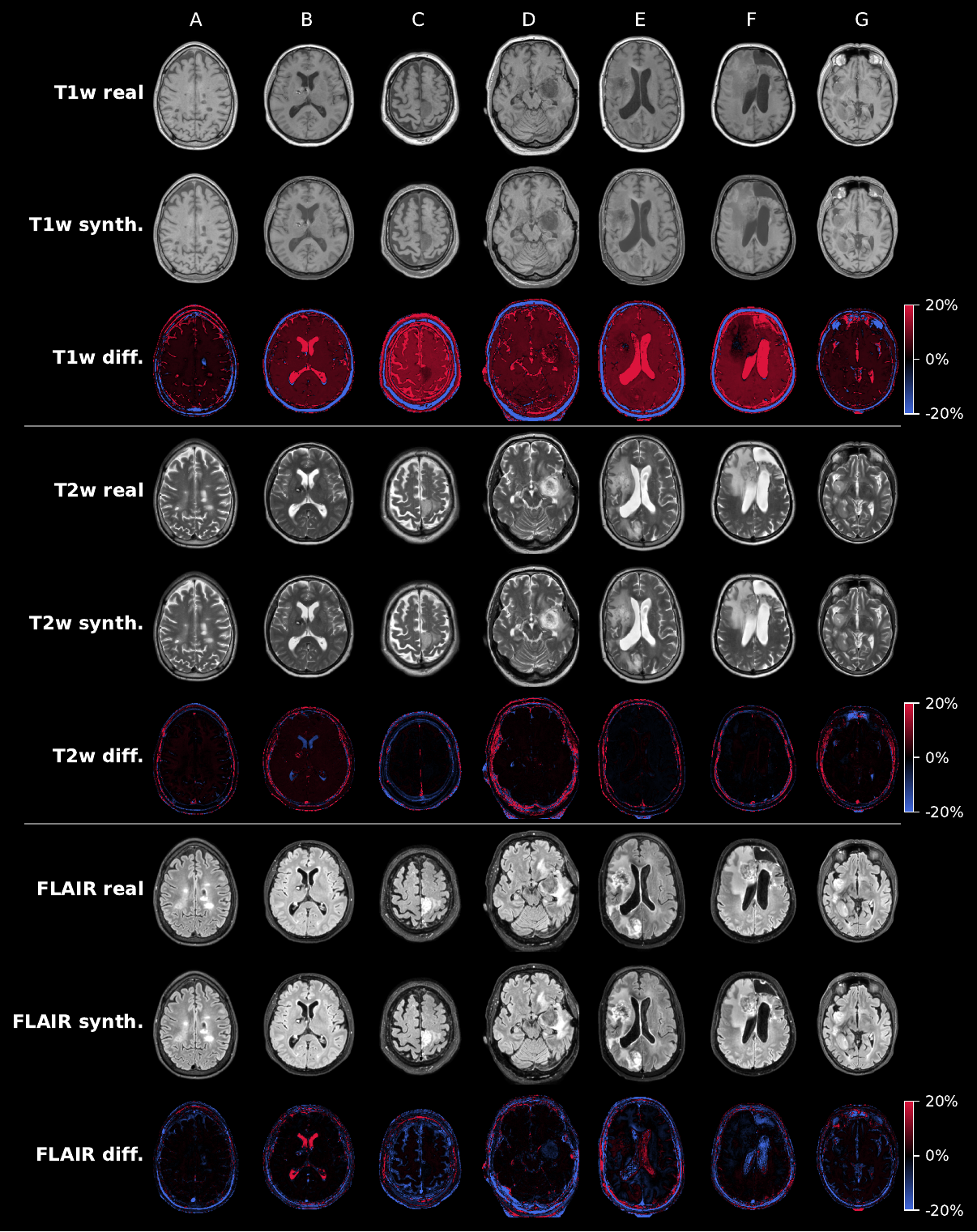}
    \caption{Comparison of real and synthesized conventional MRIs for seven representative subjects from the clinical archive test set. Generated quantitative maps of these slice are shown in the main of the paper (Figure \ref{fig:test_set_slices}). For each contrast (T1w, T2w, FLAIR), the first row shows the acquired MRI, the second row shows the corresponding MRI synthesized from the generated quantitative T1, T2, and PD maps using the Bloch-based physical signal model, and the third row shows the percentage difference between acquired and synthesized images.}
    \label{fig:app_synthetic_mri}
\end{figure}

Figure \ref{fig:app_synthetic_mri} shows conventional MRIs synthesized from the generated quantitative maps (from Figure \ref{fig:test_set_slices}) using the Bloch-based physical signal model, alongside difference maps between the acquired and synthesized images.

\clearpage

\section{Within-subject reproducibility}
\label{sec:appendix_retrospective}

Table \ref{tab:mri_params_retrospective_subjects} shows the sequence parameters used for the conventional MRIs of the included subjects for the within-subject reproducibility analysis from the clinical test set.

\begin{table}[h]
    \centering
    \caption{MRI sequence parameters of the two scan sessions (S1 and S2) for the four included subjects from the clinical test set. Parameters include repetition time (TR), echo time (TE), inversion time (TI), and flip angle (FA). Also the scanner system used for each scan session is listed. Note that S1 and S2 of subject 2--4 were acquired on different systems.}
    \label{tab:mri_params_retrospective_subjects}
    \small
    \renewcommand{\arraystretch}{1.2}
        \setlength{\tabcolsep}{3pt} 
        \begin{tabular}{@{}ccc ccc ccc cccc@{}}
            \toprule
            \multirow{2}{*}{\textbf{Subj}} & \multirow{2}{*}{\textbf{Sess}} & \multirow{2}{*}{\textbf{Scanner system}} & \multicolumn{3}{c}{\textbf{T1w: 3D Spoiled GRE}} & \multicolumn{3}{c}{\textbf{T2w: 2D TSE}} & \multicolumn{4}{c}{\textbf{T2-FLAIR: 3D TSE}} \\
            \cmidrule(lr){4-6} \cmidrule(lr){7-9} \cmidrule(lr){10-13}
             & & & \textbf{TR [ms]} & \textbf{TE [ms]} & \textbf{FA [$^\circ$]} & \textbf{TR [ms]} & \textbf{TE [ms]} & \textbf{FA [$^\circ$]} & \textbf{TR [ms]} & \textbf{TE [ms]} & \textbf{TI [ms]} & \textbf{FA [$^\circ$]} \\
            \midrule
            
            \multirow{2}{*}{1} 
            & S1 & Ingenia Elition X & 5.28 & 2.37 & 10 & 4339 & 80 & 90 & 4800 & 341 & 1650 & 90 \\
            & S2 & Ingenia Elition X & 5.26 & 2.37 & 10 & 4339 & 80 & 90 & 4800 & 338 & 1650 & 90 \\
            \midrule
            
            \multirow{2}{*}{2} 
            & S1 & Achieva           & 5.26 & 2.36 & 10 & 4275 & 80 & 90 & 4800 & 290 & 1650 & 90 \\
            & S2 & Ingenia CX        & 5.25 & 2.37 & 10 & 4276 & 80 & 90 & 4800 & 291 & 1650 & 90 \\
            \midrule
            
            \multirow{2}{*}{3} 
            & S1 & Ingenia           & 5.24 & 2.35 & 10 & 4275 & 80 & 90 & 4800 & 291 & 1650 & 90 \\
            & S2 & Ingenia Elition X & 5.29 & 2.38 & 10 & 4339 & 80 & 90 & 4800 & 332 & 1650 & 90 \\
            \midrule
            
            \multirow{2}{*}{4} 
            & S1 & Achieva           & 5.26 & 2.37 & 10 & 4339 & 80 & 90 & 4800 & 330 & 1650 & 90 \\
            & S2 & Ingenia Elition X & 5.16 & 2.30 & 10 & 4339 & 80 & 90 & 4800 & 342 & 1650 & 90 \\
            
            \bottomrule
        \end{tabular}
    \end{table}
    
Table \ref{tab:within_subject_means_compact_grid} summarizes the mean $\pm$ SD quantitative values for WM and GM across five subjects included in the subject-specific reproducibility analysis. For subjects 1--4 (included from the clinical test set), values are reported for two separate scan sessions (S1 and S2). For the prospectively acquired volunteer data, values represent a comparison between two distinct clinical protocols (P1 and P2).

\begin{table}[h]
\centering
\caption{Mean $\pm$ SD quantitative values for T1, T2, and PD. Subjects 1--4 represent scan sessions (S1, S2), while the volunteer data is acquired prospectively with 2 distinct protocols (P1, P2).}
\label{tab:within_subject_means_compact_grid}
\renewcommand{\arraystretch}{1.2}
\begin{tabular}{ll cc cc cc}
\toprule
\multirow{2}{*}{\textbf{Subject}} & \multirow{2}{*}{\textbf{S / P}} & \multicolumn{2}{c}{\textbf{T1 [ms]}} & \multicolumn{2}{c}{\textbf{T2 [ms]}} & \multicolumn{2}{c}{\textbf{PD [a.u.]}} \\
\cmidrule(lr){3-4} \cmidrule(lr){5-6} \cmidrule(lr){7-8}
& & WM & GM & WM & GM & WM & GM \\
\midrule
\multirow{2}{*}{1} & S1 & 816 $\pm$ 175 & 1163 $\pm$ 270 & 68.2 $\pm$ 5.9 & 84.1 $\pm$ 9.9 & 0.58 $\pm$ 0.08 & 0.64 $\pm$ 0.11 \\
                   & S2 & 816 $\pm$ 204 & 1161 $\pm$ 269 & 68.3 $\pm$ 6.6 & 83.1 $\pm$ 9.9 & 0.58 $\pm$ 0.09 & 0.63 $\pm$ 0.10 \\
\midrule
\multirow{2}{*}{2} & S1 & 796 $\pm$ 172 & 1125 $\pm$ 249 & 69.3 $\pm$ 6.3 & 90.1 $\pm$ 11.6 & 0.57 $\pm$ 0.08 & 0.61 $\pm$ 0.10 \\
                   & S2 & 822 $\pm$ 189 & 1152 $\pm$ 245 & 69.4 $\pm$ 5.2 & 85.8 $\pm$ 10.3 & 0.58 $\pm$ 0.09 & 0.64 $\pm$ 0.10 \\
\midrule
\multirow{2}{*}{3} & S1 & 793 $\pm$ 150 & 1193 $\pm$ 245 & 70.6 $\pm$ 6.4 & 93.8 $\pm$ 12.1 & 0.57 $\pm$ 0.07 & 0.64 $\pm$ 0.09 \\
                   & S2 & 810 $\pm$ 168 & 1198 $\pm$ 248 & 71.3 $\pm$ 7.2 & 94.4 $\pm$ 12.3 & 0.57 $\pm$ 0.08 & 0.64 $\pm$ 0.09 \\
\midrule
\multirow{2}{*}{4} & S1 & 795 $\pm$ 164 & 1138 $\pm$ 297 & 69.0 $\pm$ 5.5 & 86.3 $\pm$ 9.6 & 0.57 $\pm$ 0.08 & 0.62 $\pm$ 0.11 \\
                   & S2 & 838 $\pm$ 186 & 1178 $\pm$ 287 & 70.3 $\pm$ 5.3 & 85.0 $\pm$ 8.9 & 0.59 $\pm$ 0.08 & 0.65 $\pm$ 0.11 \\
\midrule
\midrule
\multirow{2}{*}{Volunteer} & P1 & 958 $\pm$ 126 & 1254 $\pm$ 196 & 68.7 $\pm$ 5.1 & 89.7 $\pm$ 9.6 & 0.64 $\pm$ 0.05 & 0.67 $\pm$ 0.07 \\
                   & P2 & 942 $\pm$ 144 & 1231 $\pm$ 220 & 68.5 $\pm$ 4.9 & 89.6 $\pm$ 9.5 & 0.64 $\pm$ 0.06 & 0.66 $\pm$ 0.08 \\
\bottomrule
\end{tabular}
\end{table}

\begin{figure}[h]
    \centering
    \includegraphics[width=0.9\linewidth]{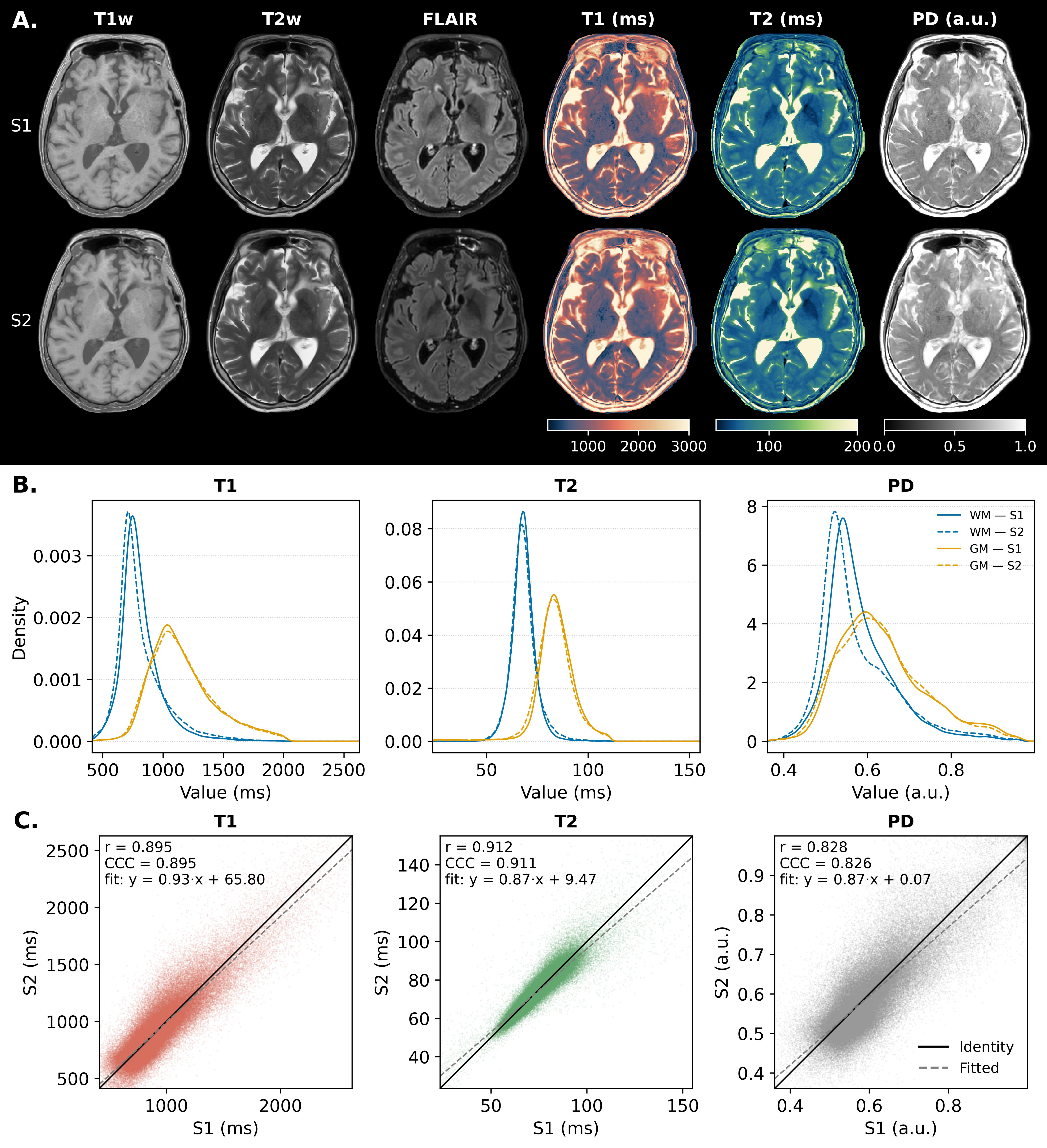}
    \caption{Subject 1: Within-subject reproducibility results for subject 1. 
            (A) Generated quantitative PD, T1, and T2 maps from the two scan sessions (S1 and S2), together with the corresponding input conventional images. The lesion is visible in the frontal region of the brain.
            (B) Whole-brain WM and GM value distributions for PD, T1, and T2. 
            (C) Voxel-wise scatterplots comparing S1 and S2 values for PD, T1, and T2 within brain parenchyma (WM + GM).}
    \label{fig:subject1}
\end{figure}
\begin{figure}[h]
    \centering
    \includegraphics[width=0.9\linewidth]{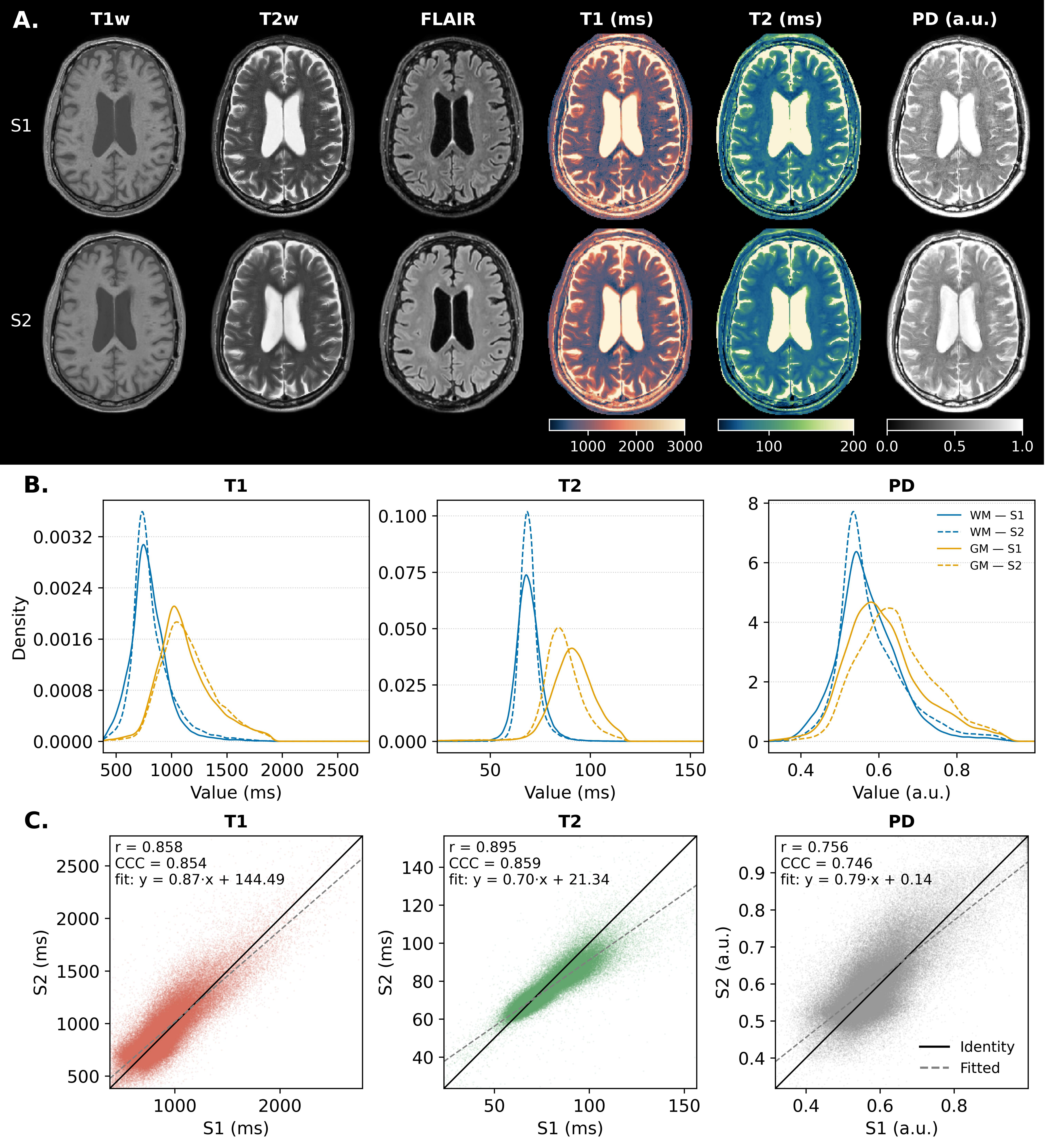}
    \caption{Subject 2: Within-subject reproducibility results for subject 2. 
            (A) Generated quantitative PD, T1, and T2 maps from the two scan sessions (S1 and S2), together with the corresponding input conventional images.
            (B) Whole-brain WM and GM value distributions for PD, T1, and T2. 
            (C) Voxel-wise scatterplots comparing S1 and S2 values for PD, T1, and T2 within brain parenchyma (WM + GM).}
    \label{fig:subject2}
\end{figure}
\begin{figure}[h]
    \centering
    \includegraphics[width=0.9\linewidth]{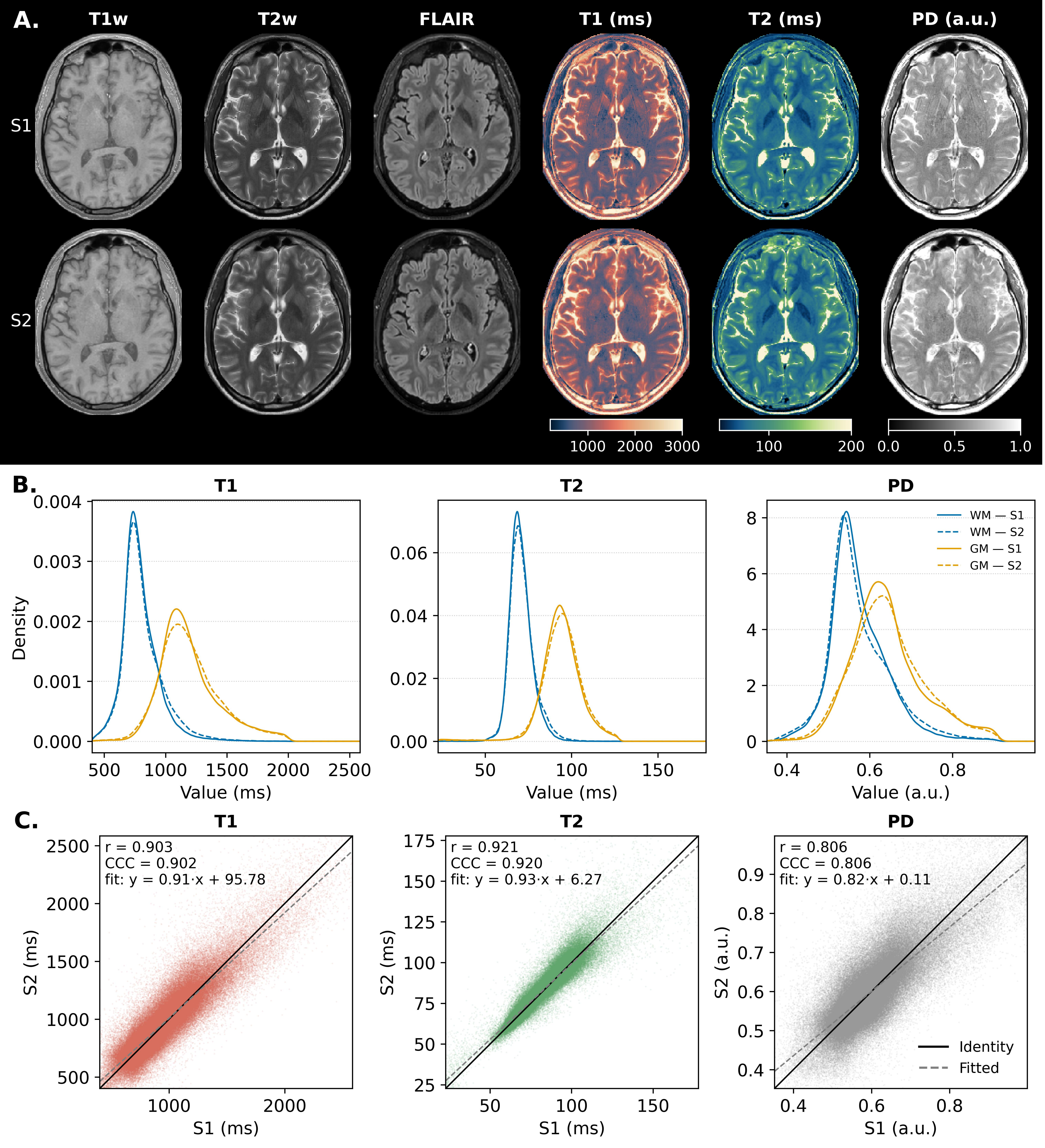}
    \caption{Subject 3: Within-subject reproducibility results for subject 3. 
            (A) Generated quantitative PD, T1, and T2 maps from the two scan sessions (S1 and S2), together with the corresponding input conventional images.
            (B) Whole-brain WM and GM value distributions for PD, T1, and T2. 
            (C) Voxel-wise scatterplots comparing S1 and S2 values for PD, T1, and T2 within brain parenchyma (WM + GM).}
    \label{fig:subject3}
\end{figure}
\begin{figure}[h]
    \centering
    \includegraphics[width=0.9\linewidth]{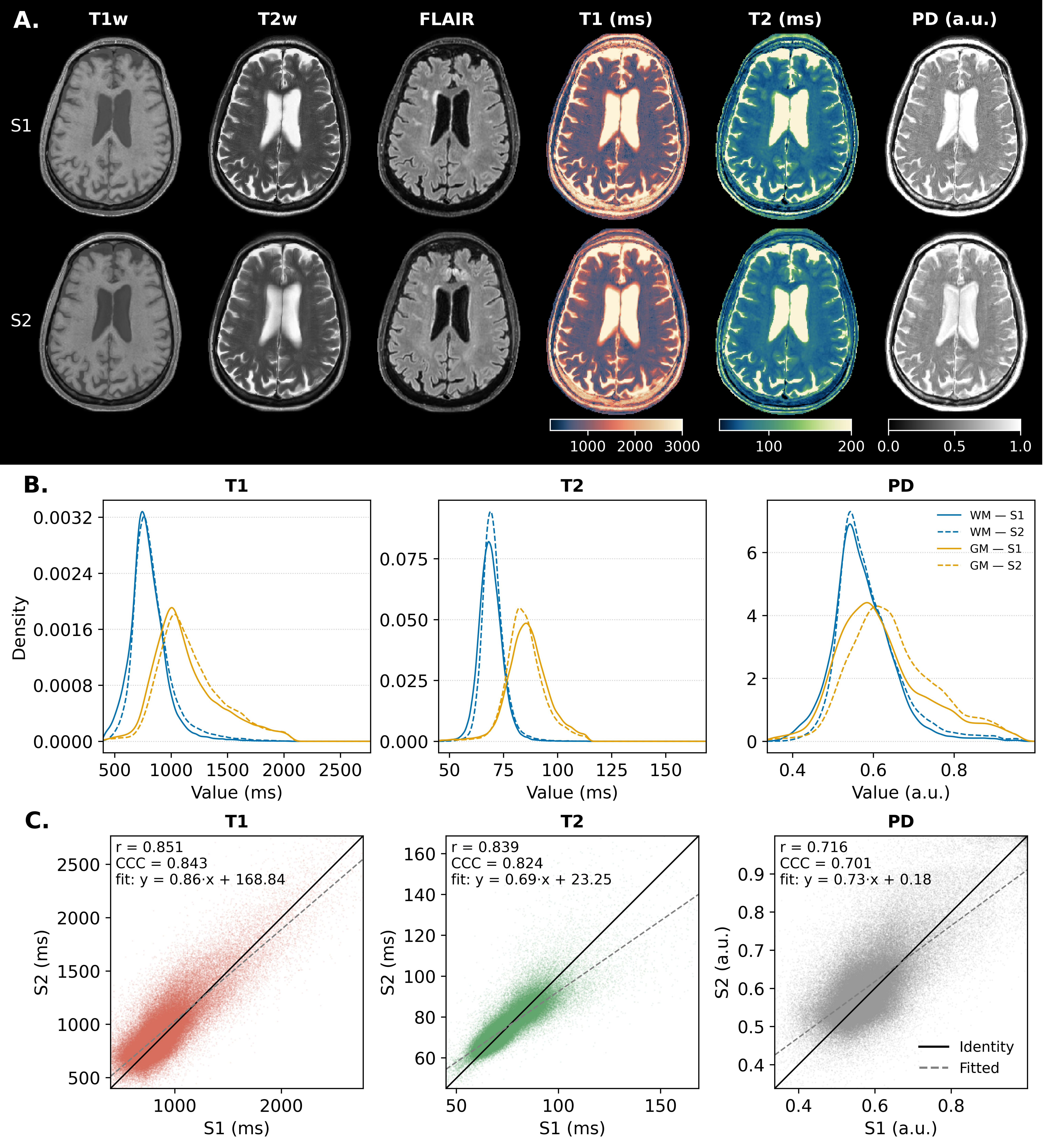}
    \caption{Subject 4: Within-subject reproducibility results for subject 4. 
            (A) Generated quantitative PD, T1, and T2 maps from the two scan sessions (S1 and S2), together with the corresponding input conventional images.
            (B) Whole-brain WM and GM value distributions for PD, T1, and T2. 
            (C) Voxel-wise scatterplots comparing S1 and S2 values for PD, T1, and T2 within brain parenchyma (WM + GM).}
    \label{fig:subject4}
\end{figure}

Figures \ref{fig:subject1}-\ref{fig:subject4} illustrate the subject-specific reproducibility results for the four included subjects from the clinical test set.

\clearpage

\section{Ablation study regularization terms and constraints}
\label{sec:appendix_ablation_study}

\begin{figure}[ht]
    \centering
    \includegraphics[width=0.90\linewidth]{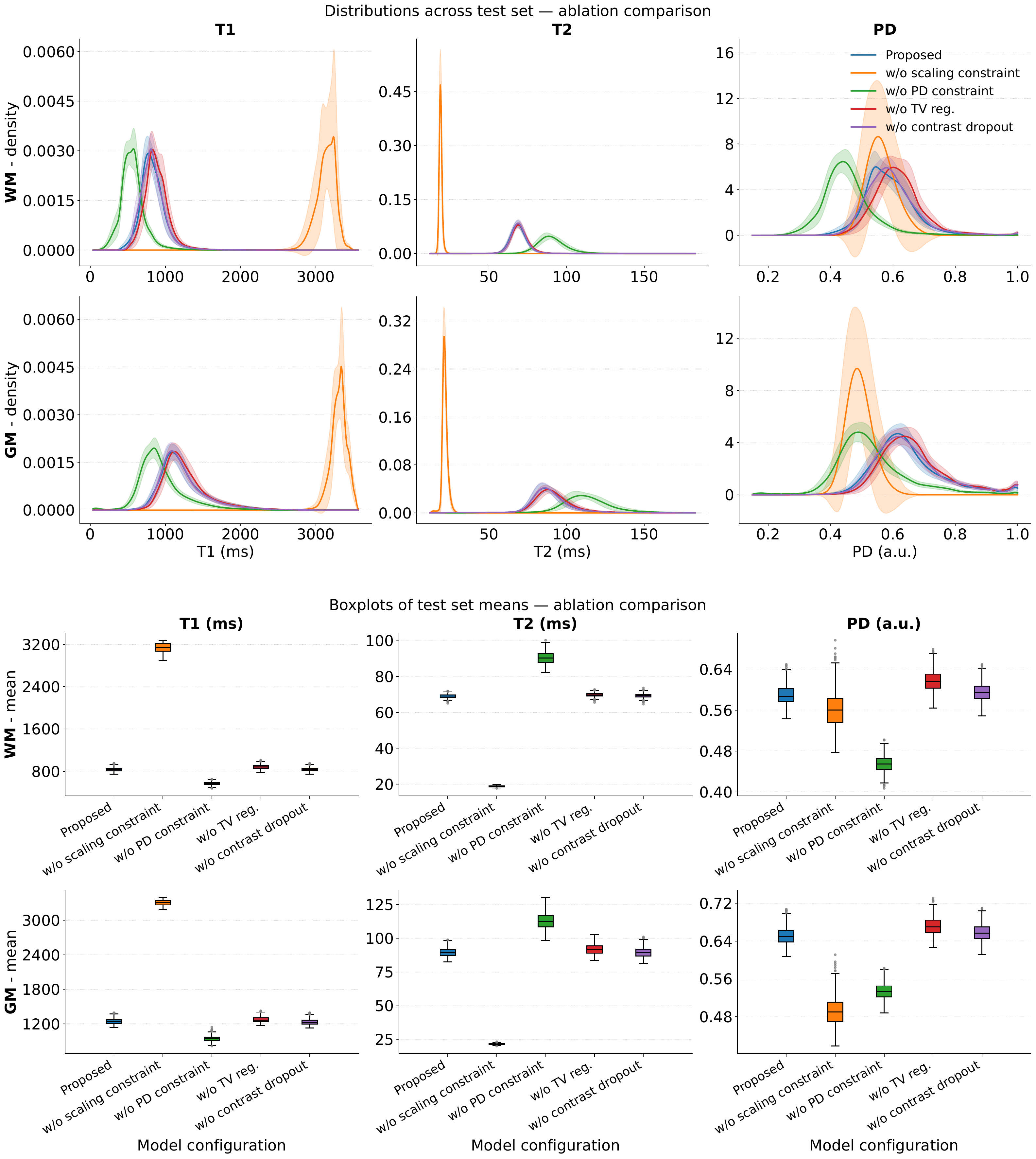}
    \caption{Summary of quantitative T1, T2, and PD values across all 603 clinical test sessions for the proposed model and the four ablated configurations. Top: distributions of T1, T2, and PD values values in white matter (WM) and gray matter (GM). Solid lines indicate the mean value across sessions and shaded areas indicate one standard deviation. Bottom: boxplots of session-wise mean WM and GM values for T1, T2, and PD. }
    \label{fig:ablation_combined_stacked}
\end{figure}

\begin{figure}[ht]
    \centering
    \includegraphics[width=0.95\linewidth]{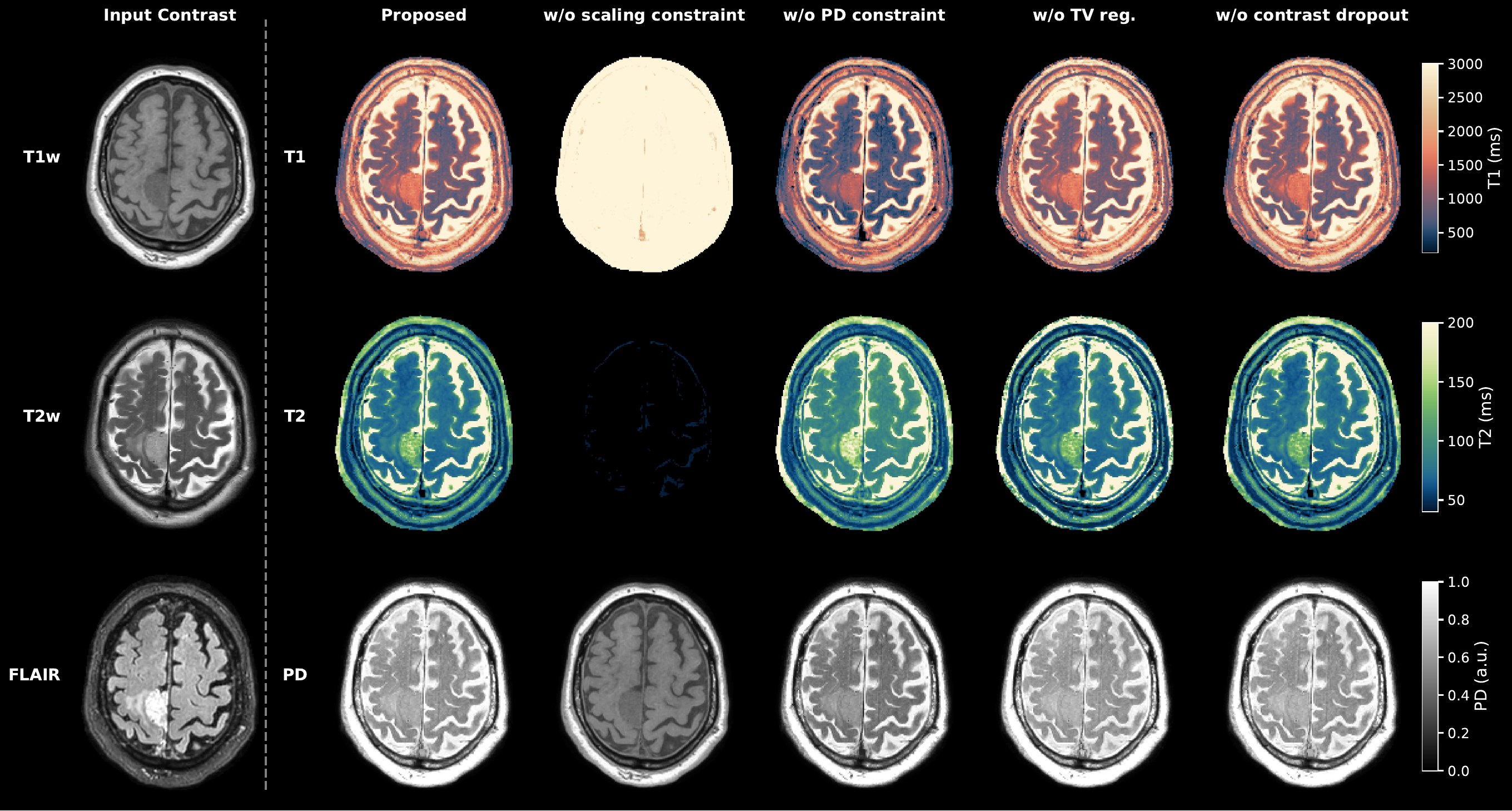}
    \caption{Representative visual comparison of the quantitative T1, T2, and PD maps generated by the proposed model and the four ablated configurations, visualizing Subject C from Figure 2 of the main manuscript.}
    \label{fig:ablation_grid}
\end{figure}

\begin{figure}[ht]
    \centering
    \includegraphics[width=0.90\linewidth]{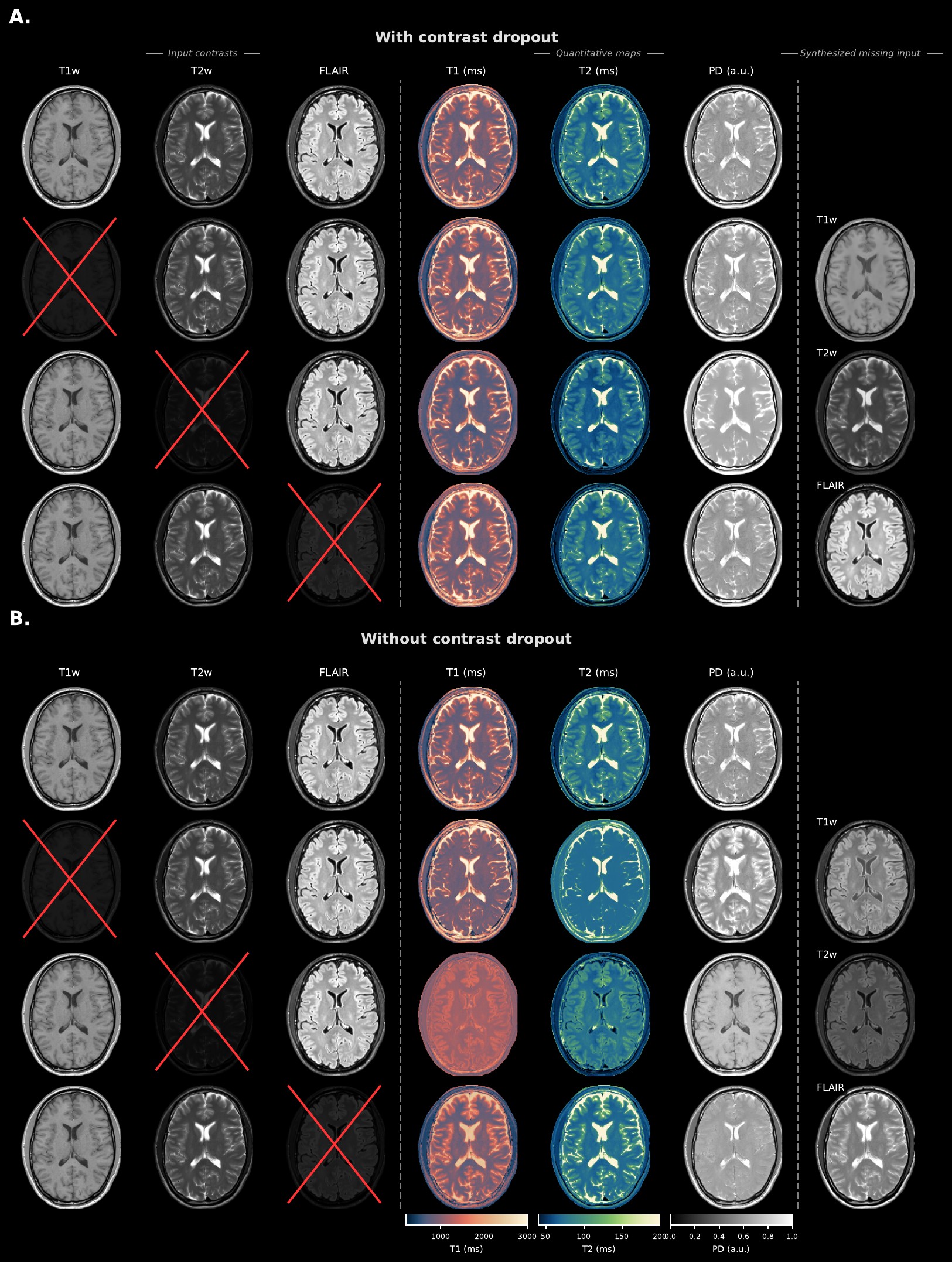}
    \caption{Visual impact of missing input modalities during inference, evaluated on scan 1 of the prospective volunteer data. \textbf{(A)} The proposed model, trained with the contrast dropout strategy. \textbf{(B)} Ablated model, trained without contrast dropout. For both models, the rows display the input contrasts (T1w, T2w, FLAIR) alongside the generated quantitative PD, T1, and T2 maps when all inputs are provided, or when the T1w, T2w, or FLAIR input is omitted (indicated by red crosses). The last column shows the synthesized missing input image, using the Bloch-based equations on the generated quantitative maps.}
    \label{fig:missing_modality}
\end{figure}

To evaluate the individual contributions of the global scaling constraints, proton-density (PD) soft constraint, total variation (TV) regularization, and contrast dropout during training, we performed a systematic ablation study. We trained separate models using identical baseline settings, omitting one constraint or regularization term at a time. To assess the impact of these components, we analyzed the models across several dimensions: overall quantitative parameter distributions of the test set (Figure \ref{fig:ablation_combined_stacked}), visual quality of the generated quantitative maps (Figure \ref{fig:ablation_grid}), voxel-wise subject-specific reproducibility (Table \ref{tab:ablation_study}), and robustness to missing input modalities (Figure \ref{fig:missing_modality}).

Removing the global scaling constraint ("w/o scaling constraint") prevents the model from generating accurate quantitative maps. Generating quantitative maps is an ill-posed problem with multiple potential solutions. Without this constraint, the network can pair inaccurate parameter estimates with arbitrary scaling factors to produce similar synthesized images. As a result, the generated maps are physically implausible (Figure \ref{fig:ablation_combined_stacked} and \ref{fig:ablation_grid}). For instance, T1 values are overestimated, and T2 values are underestimated compared to literature ranges. Additionally, the PD map seems to mimic the T1w input contrast.

Omitting the soft PD lower bound ("w/o PD constraint") results in quantitative values that deviate from literature ranges. Specifically, T1 values in WM and GM drop below expected literature ranges, while T2 values are overestimated (Figure \ref{fig:ablation_combined_stacked}).

Excluding TV regularization ("w/o TV reg.") maintains a globally stable mean distribution across the test set population (Figure \ref{fig:ablation_combined_stacked}), comparable to the full proposed model. However, Table \ref{tab:ablation_study} shows that removing this regularizer can decrease the voxel-wise reproducibility metrics (Pearson $r$ and $CCC$), particularly in the generated T2 maps.

Removing the contrast dropout strategy ("w/o contrast dropout") does not noticeably alter the generated maps when all inputs are present (Figures \ref{fig:ablation_combined_stacked} and \ref{fig:ablation_grid}). Also, compared to the full proposed model, the voxel-wise reproducibility metrics remain similar (Table \ref{tab:ablation_study}). However, while this study assumes all three conventional contrasts (T1w, T2w, and FLAIR) are available, real-world clinical data is rarely perfect. In clinical practice, a specific contrast might be omitted from a protocol or corrupted by factors such as patient motion. The necessity of the contrast dropout strategy becomes clear in these missing-input scenarios (Figure \ref{fig:missing_modality}). Without this strategy, the model loses robustness and the generated maps degrade substantially. For instance, omitting the T2w input yields incorrect CSF T1 and T2 values. Similarly, omitting the T1w input severely reduces WM/GM contrast in both the T1 and T2 maps. Conversely, the model trained with contrast dropout handles missing inputs consistently better. Furthermore, as demonstrated in the final column of Figure \ref{fig:missing_modality}, this strategy enables the network to successfully synthesize missing input contrasts. If contrast dropout is not applied during training, this synthesis fails: the inaccuracies in the quantitative maps cause the synthesized missing inputs to also be inaccurate (e.g., the synthesized FLAIR image looks more like a T2w image, while the T2w looks more like a FLAIR). While exploratory, these findings highlight a valuable future research direction. Contrast dropout not only secures model robustness against missing inputs, but also provides a framework to synthesize missing conventional MR images.

\begin{table}[htbp]
\centering
\caption{Voxel-wise within-subject reproducibility metrics across different model ablation configurations for the same subjects as in Table 4 in the main paper. Pearson correlation coefficient ($r$), concordance correlation coefficient ($CCC$), and mean relative difference are evaluated for T1, T2, and PD within the brain parenchyma (WM + GM). Vol.: prospective volunteer.}
\label{tab:ablation_study}
\renewcommand{\arraystretch}{1.1}
\setlength{\tabcolsep}{4.5pt}
\footnotesize
\begin{tabular}{l l ccc ccc ccc}
\toprule
\multirow{2}{*}{\textbf{Model}} & \multirow{2}{*}{\textbf{Subject}} &
\multicolumn{3}{c}{\textbf{T1}} & \multicolumn{3}{c}{\textbf{T2}} & \multicolumn{3}{c}{\textbf{PD}} \\
\cmidrule(lr){3-5} \cmidrule(lr){6-8} \cmidrule(lr){9-11}
& & $r$ & $CCC$ & Mean diff. & $r$ & $CCC$ & Mean diff. & $r$ & $CCC$ & Mean diff. \\
\midrule
\multirow{6}{*}{\textbf{Proposed}}
& 1 & 0.895 & 0.895 & 11.3\% & 0.912 & 0.911 & 3.9\% & 0.828 & 0.826 & 7.8\% \\
& 2 & 0.858 & 0.854 & 13.6\% & 0.895 & 0.859 & 5.6\% & 0.756 & 0.746 & 9.6\% \\
& 3 & 0.903 & 0.902 & 10.3\% & 0.921 & 0.920 & 4.1\% & 0.806 & 0.806 & 7.2\% \\
& 4 & 0.851 & 0.843 & 14.8\% & 0.839 & 0.824 & 5.1\% & 0.716 & 0.701 & 10.2\% \\
& Vol. & 0.932 & 0.928 & 6.9\% & 0.947 & 0.944 & 2.9\% & 0.858 & 0.844 & 5.1\% \\
\midrule
\multirow{6}{*}{\textbf{w/o scaling constraint}}
& 1 & 0.868 & 0.867 & 2.0\% & 0.931 & 0.926 & 2.5\% & 0.827 & 0.570 & 7.5\% \\
& 2 & 0.801 & 0.792 & 1.6\% & 0.915 & 0.910 & 2.7\% & 0.859 & 0.803 & 4.8\% \\
& 3 & 0.896 & 0.889 & 1.7\% & 0.938 & 0.937 & 2.3\% & 0.862 & 0.861 & 3.4\% \\
& 4 & 0.783 & 0.498 & 4.0\% & 0.887 & 0.857 & 3.5\% & 0.741 & 0.713 & 5.2\% \\
& Vol. & 0.899 & 0.894 & 0.7\% & 0.963 & 0.962 & 1.7\% & 0.941 & 0.940 & 2.7\% \\
\midrule
\multirow{6}{*}{\textbf{w/o PD constraint}}
& 1 & 0.893 & 0.892 & 16.5\% & 0.879 & 0.879 & 5.1\% & 0.841 & 0.837 & 9.1\% \\
& 2 & 0.857 & 0.856 & 20.5\% & 0.814 & 0.775 & 8.0\% & 0.783 & 0.780 & 11.7\% \\
& 3 & 0.899 & 0.899 & 14.6\% & 0.895 & 0.894 & 5.3\% & 0.833 & 0.832 & 9.0\% \\
& 4 & 0.853 & 0.829 & 24.9\% & 0.739 & 0.647 & 8.5\% & 0.750 & 0.708 & 13.5\% \\
& Vol. & 0.926 & 0.919 & 9.8\% & 0.947 & 0.946 & 4.0\% & 0.869 & 0.851 & 6.6\% \\
\midrule
\multirow{6}{*}{\textbf{w/o TV reg.}}
& 1 & 0.890 & 0.889 & 10.7\% & 0.868 & 0.868 & 4.0\% & 0.814 & 0.814 & 7.6\% \\
& 2 & 0.855 & 0.848 & 12.5\% & 0.894 & 0.879 & 5.5\% & 0.744 & 0.729 & 9.0\% \\
& 3 & 0.899 & 0.898 & 9.4\% & 0.451 & 0.377 & 4.9\% & 0.805 & 0.805 & 6.7\% \\
& 4 & 0.849 & 0.831 & 14.2\% & 0.742 & 0.726 & 5.3\% & 0.704 & 0.674 & 9.9\% \\
& Vol. & 0.937 & 0.932 & 6.1\% & 0.744 & 0.662 & 3.3\% & 0.863 & 0.852 & 4.7\% \\
\midrule
\multirow{6}{*}{\textbf{w/o contrast dropout}}
& 1 & 0.897 & 0.895 & 11.1\% & 0.913 & 0.911 & 3.7\% & 0.833 & 0.828 & 7.9\% \\
& 2 & 0.854 & 0.853 & 12.8\% & 0.895 & 0.863 & 5.6\% & 0.747 & 0.744 & 8.5\% \\
& 3 & 0.899 & 0.897 & 10.8\% & 0.862 & 0.855 & 4.5\% & 0.811 & 0.811 & 7.5\% \\
& 4 & 0.854 & 0.848 & 13.8\% & 0.841 & 0.824 & 5.1\% & 0.717 & 0.707 & 9.8\% \\
& Vol. & 0.931 & 0.925 & 6.9\% & 0.930 & 0.928 & 3.1\% & 0.853 & 0.841 & 5.3\% \\
\bottomrule
\end{tabular}
\end{table}

\clearpage

\section{Multiple sequence types}
\label{sec:appendix_multi_sequences}


This section presents results from experiments in which multiple sequence types for the same contrast weighting were combined during both training and evaluation of our framework. For this analysis, we constructed an additional clinical dataset beyond the one used in the main paper. For this supplementary dataset, we expanded the inclusion window to 2015–2023. We incorporated additional $3$~T MRI sessions containing T1w 2D SE, T1w 3D Gradient Echo with preparation pulse (IR-GRE), and T2-FLAIR 2D TSE sequences (Table \ref{tab:supp_clinical_archive_info}). This dataset contains a total of 5,531 scan sessions. This allowed us to assess the framework's performance under heterogeneity in sequence types.

We split this supplementary dataset into training, validation, and test sets in an 80/5/15\% ratio. The framework is trained in the same fashion as described in the main paper. Since scan sessions containing T1w 2D SE and 2D T2-FLAIR TSE were underrepresented in the dataset, we oversampled them so that they appeared five times more frequently than their original frequency during training. The same signal models were used as in the main paper, with the addition of the following Bloch-based signal model for the T1w 3D IR-GRE sequence (Equation \ref{eq:ir-gre}).

\begin{equation}
    \label{eq:ir-gre}
    S_{\text{IR-GRE}} = PD \cdot \frac{1 - 2\exp(-TI/T1) + \exp(-TR/T1)}{1 + \cos(FA) \cdot \exp(-TR/T1)} \cdot \sin(FA) \cdot \exp(-TE/T2)
\end{equation}

\begin{table}[h]
\centering
\renewcommand{\arraystretch}{1.2}
\caption{Sequence parameters of the conventional conventional brain MRI scans in the supplementary clinical dataset. Each contrast (T1w, T2w, FLAIR) includes one or more sequence types. For each sequence type the repetition time (TR), echo time (TE), inversion time (TI), and flip angle (FA) are listed. SE: Spin Echo; Spoiled GRE: Spoiled Gradient-Echo; IR-GRE: Gradient Echo with preparation pulse; TSE: Turbo Spin Echo.}
\label{tab:supp_clinical_archive_info}

\begin{tabular}{rcccccc}
\toprule
& \multicolumn{3}{c}{\textbf{T1w}} 
& \multicolumn{1}{c}{\textbf{T2w}} 
& \multicolumn{2}{c}{\textbf{T2-FLAIR}} \\

\cmidrule(lr){2-4} \cmidrule(lr){5-5} \cmidrule(lr){6-7}
& \textbf{3D Spoiled GRE} & \textbf{3D IR-GRE} & \textbf{2D SE}
& \textbf{2D TSE}
& \textbf{3D TSE} & \textbf{2D TSE} \\

\midrule
\textbf{N} 
& 4,121 & 1,152 & 258 
& 5,531 
& 5,267 & 264 \\

\textbf{Gadolinium} 
& No & No & No
& Yes (91.4\%)
& Yes (28.3\%) & No \\

\textbf{TR [ms]} 
& 5.03 -- 5.40 & 2500 & 475 -- 550
& 2780 -- 4557
& 4800 & 10000 -- 11000 \\

\textbf{TE [ms]} 
& 2.247 -- 2.493 & 3.53 -- 4.26 & 13
& 80
& 275 -- 370  & 120 -- 125 \\

\textbf{TI [ms]} 
& NA & 1004
& NA
& NA
& 1650 & 2800 \\

\textbf{FA [$^\circ$]} 
& 10 & 8 & 70 or 90
& 90
& 90 & 90 \\

\bottomrule
\end{tabular}
\end{table}

To validate this trained model, we additionally collected prospective data. A healthy volunteer was scanned three times using different protocols, each employing different sequence types and parameters for the various contrast weightings, similar to those included in the supplementary clinical dataset. The used sequences and sequence parameters are detailed in Table \ref{tab:prospective_acquisitions_info_supp}. Using the trained model we generated quantitative maps from these three different input protocols and compared them visually.

\begin{table}[h]
\centering
\renewcommand{\arraystretch}{1.2}
\caption{Sequence types and parameters for the three different protocols (P1, P2, and P3) of the prospectively acquired volunteer data. Sequence types are specified for each protocol.}
\label{tab:prospective_acquisitions_info_supp}

\setlength{\tabcolsep}{3pt} 
\begin{tabular}{rccccccccc}
\toprule
 & \multicolumn{3}{c}{\textbf{T1w}} 
 & \multicolumn{3}{c}{\textbf{T2w}}
 & \multicolumn{3}{c}{\textbf{T2-FLAIR}} \\
\cmidrule(lr){2-4}\cmidrule(lr){5-7}\cmidrule(lr){8-10}
 & P1 & P2 & P3
 & P1 & P2 & P3
 & P1 & P2 & P3 \\
\midrule
\textbf{Sequence}         & 3D Spoiled GRE & 3D IR-GRE & 2D SE & 2D TSE & 2D TSE & 2D TSE & 3D TSE & 3D TSE & 2D TSE \\
\textbf{Gadolinium}       & No & No & No & No & No & No & No & No & No \\
\textbf{TR [ms]}          & 5.07 & 2500 & 720 & 4340 & 4340 & 4990 & 4800 & 4800 & 10000 \\
\textbf{TE [ms]}          & 2.273 & 3.794 & 14 & 80 & 80 & 80 & 336 & 335 & 120 \\
\textbf{TI [ms]}          & NA & 1004 & NA & NA & NA & NA & 1650 & 1650 & 2800 \\
\textbf{FA [$^\circ$]}    & 10 & 8 & 70 & 90 & 90 & 90 & 90 & 90 & 90 \\
\bottomrule
\end{tabular}
\end{table}

\begin{figure}[h]
    \centering
    \includegraphics[width=0.95\linewidth]{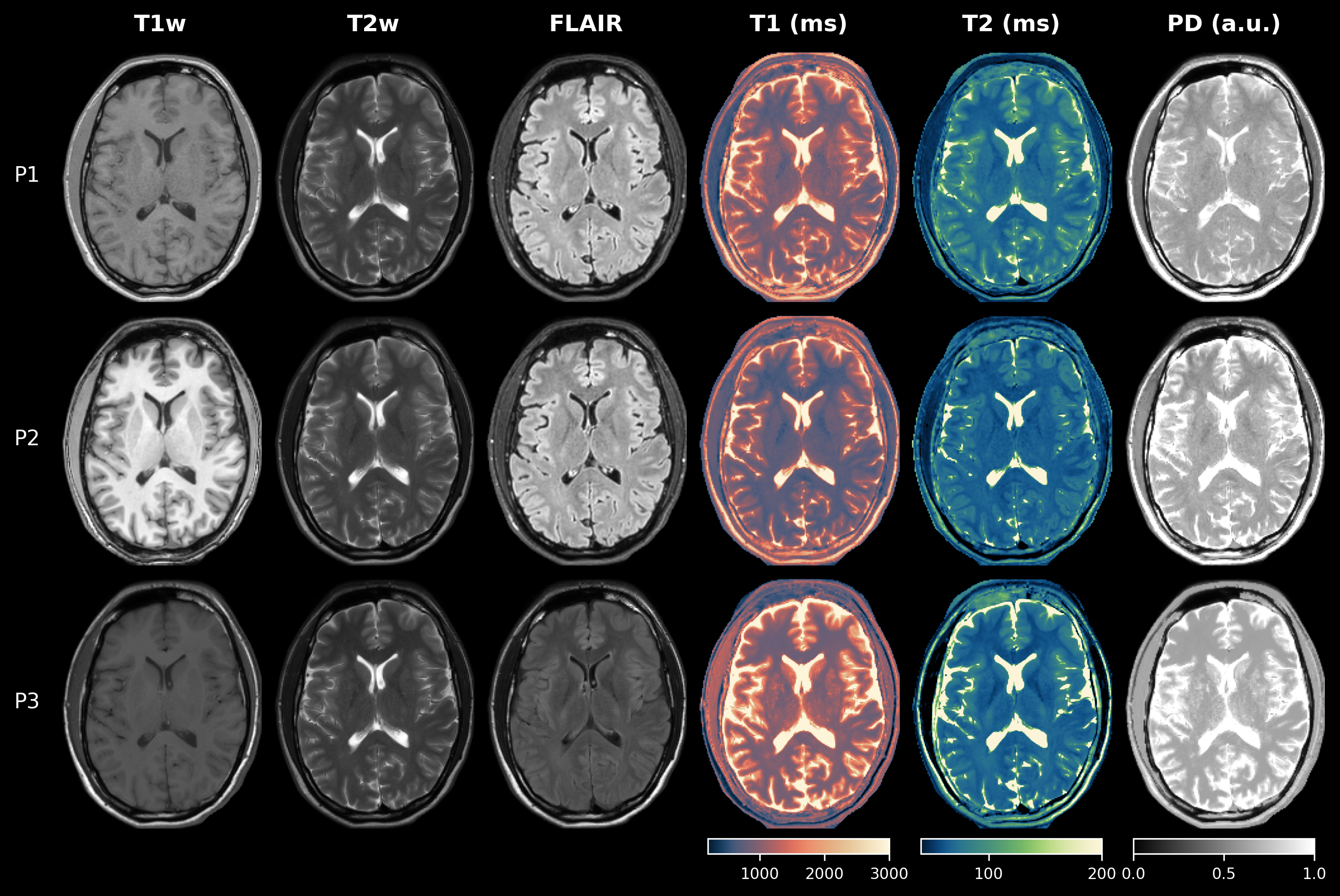}
    \caption{Representative quantitative T1, T2, and PD maps generated from different input protocols in a healthy volunteer. Each row shows the conventional MRI inputs (T1w, T2w, FLAIR) and the corresponding generated quantitative maps. The maps appear visually inconsistent across protocols, demonstrating variability in the outputs depending on the input sequence type.}
    \label{fig:figure_prospective_supp}
\end{figure}

Figure \ref{fig:figure_prospective_supp} shows the resulting quantitative maps, together with the corresponding conventional MRI inputs. Across the different input protocols (each using different sequence types) the generated T1, T2, and PD maps appear visually inconsistent. This indicates that the outputs are not reproducible within the subject when the input sequence type varies.

\end{document}